\documentclass[pdflatex,sn-nature]{sn-jnl}

\usepackage{graphicx}%
\usepackage{multirow}%
\usepackage{amsmath,amssymb,amsfonts}%
\usepackage{amsthm}%
\usepackage[title]{appendix}%
\usepackage{xcolor}%
\usepackage{array}%
\usepackage{textcomp}%
\usepackage{manyfoot}%
\usepackage{booktabs}%
\usepackage{algorithm}%
\usepackage{algorithmicx}%
\usepackage{algpseudocode}%
\usepackage{listings}%
\usepackage{bbm}
\usepackage{lineno}
\usepackage{subcaption}
\usepackage{gensymb}
\usepackage{fix-cm}

\begin{document}

\title[DiffSR]{Diffusion-Based Super-Resolution of Adriatic Sea Oceanographic Fields}

\author[]{\fnm{Rajat} \sur{Srivastava}}\email{rajat.srivastava@unisalento.it}
\author[]{\fnm{Muhammad} \sur{Sarmad}}\email{muhammad.sarmad@unisalento.it}
\author[]{\fnm{Emanuele} \sur{Mele}}\email{emanuele.mele@unisalento.it}
\author[]{\fnm{Massimo} \sur{Cafaro}}\email{massimo.cafaro@unisalento.it}
\author[]{\fnm{Marco} \sur{Pulimeno}}\email{marco.pulimeno@unisalento.it}
\author*[ ]{\fnm{Italo} \sur{Epicoco}}\email{italo.epicoco@unisalento.it}

\affil[]{\orgdiv{Department Engineering for Innovation}, \orgname{University of Salento}, \orgaddress{\street{via per Monteroni}, \city{Lecce}, \postcode{73100}, \country{Italy}}}

\abstract{
High-resolution oceanographic fields are critical for resolving mesoscale and sub-mesoscale coastal dynamics, yet their generation remains constrained by both computational cost and observational sparsity. We present \textit{OcDiffSR}, a conditional denoising diffusion probabilistic model (DDPM) for oceanographic super-resolution that reconstructs high-resolution sea-surface fields from coarse-resolution reanalysis inputs. The model is trained on ten years (2011--2020) of paired low-resolution (GLORYS12V1, 1/12\degree) and high-resolution (Mediterranean Sea Physics Reanalysis, Med~MFC, 1/24\degree) data, and evaluated on an independent test year (2009) over the Adriatic Sea. OcDiffSR employs a conditional U-Net augmented with multi-scale low-resolution encoders, cross-attention bottleneck layers, and sinusoidal seasonal embeddings via Feature-wise Linear Modulation (FiLM), enabling joint super-resolution of sea-surface temperature (SST), salinity (SSS), and horizontal velocity components with visually coherent circulation patterns. Benchmarked against bilinear interpolation and the state-of-the-art residual diffusion model CorrDiff, OcDiffSR achieves substantially lower reconstruction errors for scalar fields (RMSE$_{\mathrm{SST}}$=0.477\degree C, RMSE$_{\mathrm{SSS}}$=0.346~psu), near-unity Pearson correlation (PCC $\geq$ 0.999), and high structural similarity (SSIM $\geq$ 0.964). For dynamical vector fields, OcDiffSR outperforms both baselines in absolute error and spatial coherence, though moderate correlation (PCC = 0.64) reflects the intrinsic stochasticity of oceanic velocity fields. Daily and monthly evaluations confirm temporal robustness across all seasons. These results establish OcDiffSR as a reliable framework for high-fidelity oceanographic downscaling and reanalysis enhancement, producing fields that are visually consistent with known ocean dynamics.
}

\keywords{Denoising diffusion probabilistic models, oceanographic super-resolution, statistical downscaling, Mediterranean Sea, Adriatic Sea, reanalysis enhancement, generative deep learning, coastal ocean dynamics}

\maketitle

\section{Introduction}\label{introduction}
Advances in data-driven machine learning and artificial intelligence have substantially improved our ability to characterise complex ocean dynamics such as mesoscale eddies, submesoscale filaments, fronts, and boundary currents, with broad implications for civilian and scientific applications \cite{Hong_2025}. Ocean dynamics govern the transport of heat, salt, and nutrients, exerting direct influence on the marine ecosystem, biogeochemical cycles, and regional climate variability. To predict such complex phenomena, data-driven machine learning tools and AI techniques require high-resolution oceanographic datasets. Generating such high-resolution datasets accurately is therefore critical for operational oceanography, climate modelling, and coastal management. However, global reanalysis products are typically constrained to coarse spatial resolutions owing to the prohibitive computational cost of high-resolution ocean modelling. While coarse-resolution reanalysis products adequately represent large-scale circulation patterns, they systematically fail to resolve fine-scale structures. These small-scale structural patterns are crucial for understanding regions such as the Adriatic Sea, which exhibit strong spatial heterogeneity, complex bathymetry, and dynamically rich circulation. 

Oceanic processes exhibit pronounced multiscale and nonlinear behavior,  spanning a wide range of spatial and temporal scales and involving high-dimensional, coupled state variables. As a result, obtaining high-resolution representations of oceanic parameters remains a significant challenge for both observational systems and numerical models. Machine learning approaches, particularly generative frameworks such as diffusion models, offer a powerful paradigm for super-resolving oceanic fields by capturing complex spatial dependencies and quantifying uncertainty across scales.

To obtain a high-resolution oceanographic dataset, super-resolution techniques must be applied. The conventional methods for downscaling includes bilinear, bicubic, and spline-based interpolation methods that are simple and computationally inexpensive \cite{Han2013}. However, these techniques cannot capture complex non-linear ocean dynamics, high-frequency variability and fine spatial structures inherent in oceanic systems \cite{app15095013}. Variational and statistical downscaling methods can improve accuracy but typically require domain-specific assumptions about physical processes and often struggle to generalize to unseen dynamical regimes. 
Dynamical downscaling represents a more physically grounded alternative, whereby a high-resolution regional ocean model is nested within, or forced at its lateral boundaries by, a coarser global reanalysis or climate projection \cite{Adloff_2015, Mourre_2019}. In the oceanographic context, regional configurations of models such as NEMO \cite{Madec_2023} and ROMS \cite{Shchepetkin_2005} have been widely employed to dynamically downscale global reanalysis products over semi-enclosed basins, including the Mediterranean and Adriatic Seas \cite{Adloff_2015, Escudier_2021}. By explicitly resolving sub-grid processes through the numerical integration of the governing primitive equations, dynamical downscaling can faithfully represent mesoscale eddies, coastally trapped waves, and density-driven circulation features that are absent in coarse global products \cite{Escudier_2021, Mourre_2019}. Nevertheless, dynamical downscaling carries substantial computational costs, as the integration of high-resolution regional models requires significant HPC resources and introduces sensitivity to lateral boundary conditions, bathymetric representation, and atmospheric forcing fields \cite{Adloff_2015}. Furthermore, systematic biases inherited from the driving global model can propagate into the regional solution, limiting the accuracy of downscaled fields particularly in dynamically complex regions such as the Adriatic Sea \cite{Mourre_2019, Escudier_2021}. These limitations motivate the development of complementary data-driven approaches capable of delivering high-resolution oceanographic fields at a fraction of the computational cost, while retaining the physical coherence achieved by dynamical methods.

Generative modeling has become a significant and advancing area of research in recent years. Key model types, including generative adversarial networks (GANs) \cite{goodfellow2014gan}, variational autoencoders (VAEs) \cite{kingma2014adam}, autoregressive models \cite{oord2016pixelrecurrentneuralnetworks}, flow models \cite{dinh2017sharpminimageneralizedeep}\cite{kingma2018glowgenerativeflowinvertible}, and diffusion models \cite{sohl2015deep}\cite{Ho2020}, have driven substantial advances in generative modelling. These models have been successfully applied to a variety of tasks, such as realistic image generation \cite{brock2019largescalegantraining}\cite{razavi2019generatingdiversehighfidelityimages}\cite{esser2021tamingtransformershighresolutionimage}, image super-resolution \cite{Ledig_2017}\cite{bellkligler2020blindsuperresolutionkernelestimation}\cite{Saharia_2022}, image editing \cite{Park_2019}\cite{Huang_2022}, and text-to-image generation \cite{saharia2022photorealistictexttoimagediffusionmodels}\cite{ramesh2022hierarchicaltextconditionalimagegeneration}\cite{nichol2021improved}. Among these, denoising diffusion probabilistic models (DDPMs) have emerged as particularly compelling, providing a principled probabilistic framework for high-fidelity data reconstruction. In contrast to standard convolutional neural networks (CNNs), diffusion models iteratively remove noise from a perturbed sample, learning the underlying data distribution through a stochastic forward-reverse process. The forward diffusion process gradually introduces Gaussian noise in the input data, while the reverse process removes the noise and reconstructs the original field using a U-Net. Conditional extensions of DDPMs allow auxiliary inputs, such as coarse-resolution ocean state fields, to guide the reverse process and improve the reconstruction fidelity. These characteristics make diffusion models well-suited for reconstructing the highly nonlinear, multi-scale features of a high-resolution physics-informed dataset. Prior to the diffusion era, foundational CNN-based super-resolution methods shaped the field. SRCNN pioneered the use of convolutional neural networks for single-image super-resolution, establishing the base for learned upscaling methods \cite{dong2016srcnn}. Building on this, EDSR demonstrated that deep residual networks could further improve reconstruction quality by enabling very deep architectures without the vanishing-gradient degradation associated with very deep networks \cite{lim2017edsr}. ESRGAN later leveraged adversarial training to produce sharper and more realistic high-resolution images, highlighting the benefit of perceptual losses compared to purely pixel-wise optimization \cite{wang2018esrgan}. These works collectively provide the conceptual basis for modern diffusion-based super-resolution models. 
Beyond the general super-resolution and diffusion-model literature discussed above, several studies have specifically targeted ocean super-resolution and downscaling. \cite{10.2166/wcc.2022.291} compared CNN- and GAN-based single-image super-resolution architectures (SRCNN, ESRGAN, RRDBNet) for sea surface temperature reconstruction, while \cite{fanelli2024medsst} applied a dilated multi-scale convolutional network to super-resolve Mediterranean Sea surface temperature fields, the same basin considered in this study. Closer to the present testbed, \cite{adobbati2025adriatic} proposed a UNet-based deep learning approach for coastal downscaling of the northern Adriatic Sea, incorporating river-forcing information to reconstruct physical and ecosystem variables. For vector fields, \cite{yuan2024sshcurrents} used a convolutional residual network to downscale sea surface height and depth-averaged currents in a coastal domain. Most closely related in methodology, \cite{han2024kuroshiodiffusion} introduced a conditional generative diffusion model to downscale observed sea surface height over the Kuroshio Extension, demonstrating the applicability of diffusion-based generative modelling to ocean fields. These prior studies each target a single scalar or vector variable in isolation, and, with the exception of \cite{han2024kuroshiodiffusion}, rely on deterministic CNN/GAN architectures rather than a probabilistic diffusion formulation. To the best of our knowledge, no prior work has applied a diffusion-based model to jointly reconstruct multiple coupled scalar and vector oceanographic fields, nor has a diffusion-based approach previously been applied to the Adriatic Sea; OcDiffSR addresses both gaps.

The present study proposes a conditional denoising diffusion probabilistic model for oceanographic super-resolution (OcDiffSR). The approach overcomes the limitations of traditional interpolation and conventional CNN-based methods by exploiting the iterative and probabilistic nature of diffusion models. Conditional diffusion models, when properly configured, provide improved fidelity in high-resolution reconstructions of complex oceanographic variables \cite{sarmad2025hyperparameters}. OcDiffSR captures small-scale spatial structures and sub-mesoscale variability that are systematically lost at coarser scales. The approach also preserves circulation patterns in both scalar and vector fields that are visually consistent with the underlying ocean dynamics, though a rigorous quantitative assessment of physical consistency has not been performed in this study. In addition, the probabilistic generative formulation of OcDiffSR natively supports ensemble-based uncertainty quantification: independent high-resolution realisations can be obtained by repeated sampling of the reverse diffusion process from distinct Gaussian noise initialisations, conditioned on the same low-resolution input. This capability provides an inherent mechanism for estimating reconstruction uncertainty, that is a feature absent from deterministic interpolation and CNN-based approaches and critical for operational oceanography applications such as data assimilation and model validation. A full empirical characterisation of the ensemble spread and its calibration against reanalysis variability is left as a subject of future investigation. Furthermore, these models can also incorporate temporal and seasonal data through sinusoidal embeddings and conditional inputs, enabling more accurate and context-aware reconstructions that reflect the dynamic and cyclical nature of oceanographic processes.

The OcDiffSR model uses high-resolution reanalysis data (Mediterranean Sea Physics Reanalysis, Med MFC) \cite{CMEMS_MedMFC_PHY} as ground truth and coarse-resolution Global Sea Physics Reanalysis Data (GLORYS12V1) \cite{CMEMS_GLORYS12V1} as conditional inputs. The conditional U-Net architecture incorporates residual blocks, multi-scale conditioning, attention modules, and temporal embeddings, including sinusoidal month representations, to model seasonal variability. The model reconstructs sea-surface temperature, salinity, and zonal and meridional velocities, learning to capture fine-scale structures while preserving global coherence.
Beyond the scientific contributions, OcDiffSR is designed as a modular and reproducible computational framework, addressing the informatics requirements of large-scale geoscientific data processing. The data pipeline is implemented in PyTorch with \texttt{xarray}-based \cite{hoyer2017xarray} NetCDF-4 interfacing, supporting efficient on-demand, memory-resident access to paired multi-resolution reanalysis datasets without requiring full dataset pre-loading. The preprocessing chain, comprising temporal alignment, two-step z-score and min-max normalisation, land-sea masking, and sinusoidal temporal embedding, is fully parameterised and directly portable to other regional ocean domains and variable configurations. The distributed training configuration, deployed across four NVIDIA A100 GPUs on the Leonardo supercomputer at CINECA using the NCCL communication backend, demonstrates the scalability of conditional diffusion training for geoscientific applications at decade-scale reanalysis resolution. Together, these design choices position OcDiffSR as a transferable informatics tool for high-resolution oceanographic reconstruction, extendable to other semi-enclosed basins and, prospectively, to three-dimensional subsurface field estimation.

The present work makes five distinct contributions to the field of data-driven oceanographic downscaling. First, OcDiffSR introduces a joint conditional super-resolution framework for multiple coupled oceanographic state variables (sea surface temperature, salinity, and both horizontal velocity components) within a single end-to-end probabilistic model, thereby preserving dynamical consistency across scalar and vector fields rather than treating each variable independently. 
Second, the architecture incorporates a dedicated multi-scale low-resolution encoder that produces conditioning feature maps at each of the four resolution levels of the main U-Net, ensuring that coarse-scale physical context is propagated across all spatial frequencies of the reconstruction.
Third, temporal and seasonal variability is explicitly encoded through sinusoidal month embeddings injected via Feature-wise Linear Modulation (FiLM), enabling the model to adapt its reconstructions to the cyclical dynamics of ocean circulation and stratification without requiring explicit physical constraints. 
Fourth, the probabilistic generative formulation of the model enables ensemble-based uncertainty quantification of the reconstructed fields through repeated stochastic sampling, a capability absent from deterministic CNN-based or interpolation methods yet critical for operational oceanography applications such as data assimilation and model validation. 
Fifth, the framework is evaluated on a demanding and scientifically relevant testbed: the Adriatic Sea, a semi-enclosed basin characterised by complex bathymetry, sharp frontal gradients, and active mesoscale dynamics, using a test period (2009) that strictly predates the training window (2011--2020), thereby precluding any temporal leakage from autocorrelated reanalysis fields.

\section{Background}
\subsection{Denoising Diffusion Probabilistic Model (DDPM)}
Denoising Diffusion Probabilistic Models (DDPMs) \cite{Ho2020} represent a class of latent variable generative models that learn a data distribution by modelling the reversal of a gradual noise-corruption process. The framework comprises two complementary Markov chains: a forward process, which gradually corrupts a clean data sample with Gaussian noise, and a reverse process, which learns to recover the clean sample by iteratively removing that noise.

In the forward process, a clean data sample $x_0$ is progressively corrupted into near-isotropic Gaussian noise over $T$ discrete steps. At each timestep $t \in \{1, \ldots, T\}$, noise is injected according to the Gaussian transition:
\begin{equation}
    q(\mathbf{x}_t | \mathbf{x}_{t-1})= \mathcal{N}(\sqrt{1-\beta_t}\mathbf{x}_{t-1}, \beta_t \mathbf{I}),
\end{equation}
where $q$ denotes the forward transition distribution, $\beta_t$ is the noise variance at step $t$, and $\mathbf{I}$ is the identity matrix. For sufficiently large $T$, $\mathbf{x}_T$ approximates a sample from the isotropic Gaussian $\mathcal{N}(\mathbf{0}, \mathbf{I})$.

The reverse process models the time-reversal of the forward chain, iteratively denoising $\mathbf{x}_T$ to recover a clean sample $\mathbf{x}_0$. Each reverse step is modelled as a parameterized Gaussian transition:
\begin{equation}
    p_{\boldsymbol{\theta}}(\mathbf{x}_{t-1}|\mathbf{x}_t) = \mathcal{N}(\mu_{\boldsymbol{\theta}}(\mathbf{x}_t, t), \sigma_t^2\mathbf{I}),
\end{equation}

where $\boldsymbol{\mu}_{\boldsymbol{\theta}}(\mathbf{x}_t, t)$ is the mean predicted by a neural network and $\sigma_t^2$ is a fixed scalar variance, typically set to $\beta_t$ or to the posterior variance $\tilde{\beta}_t = \frac{1 - \bar{\alpha}_{t-1}}{1 - \bar{\alpha}_t}\beta_t$ \cite{Ho2020}. The network is trained to predict the noise $\boldsymbol{\epsilon}_{\boldsymbol{\theta}}$ injected at step $t$, by minimising the simplified variational objective:
\begin{equation}
 \mathcal{L}(\boldsymbol{\theta})=\mathbb{E}_{\mathbf{x}_{0},t,\boldsymbol{\epsilon}} \left\| \boldsymbol{\epsilon}-\boldsymbol{\epsilon_\theta} (\mathbf{x}_t,t) \right\|^2
\end{equation}

This objective admits stable gradient-based optimisation and, at inference time, high-quality samples are generated by initialising $\mathbf{x}_T \sim \mathcal{N}(\mathbf{0}, \mathbf{I})$ and iteratively applying the learned reverse transitions over $T$ steps.

\section{Methodology}
\subsection{Study Area}
\begin{figure}[ht]
    \centering
    \includegraphics[width=0.8\textwidth]{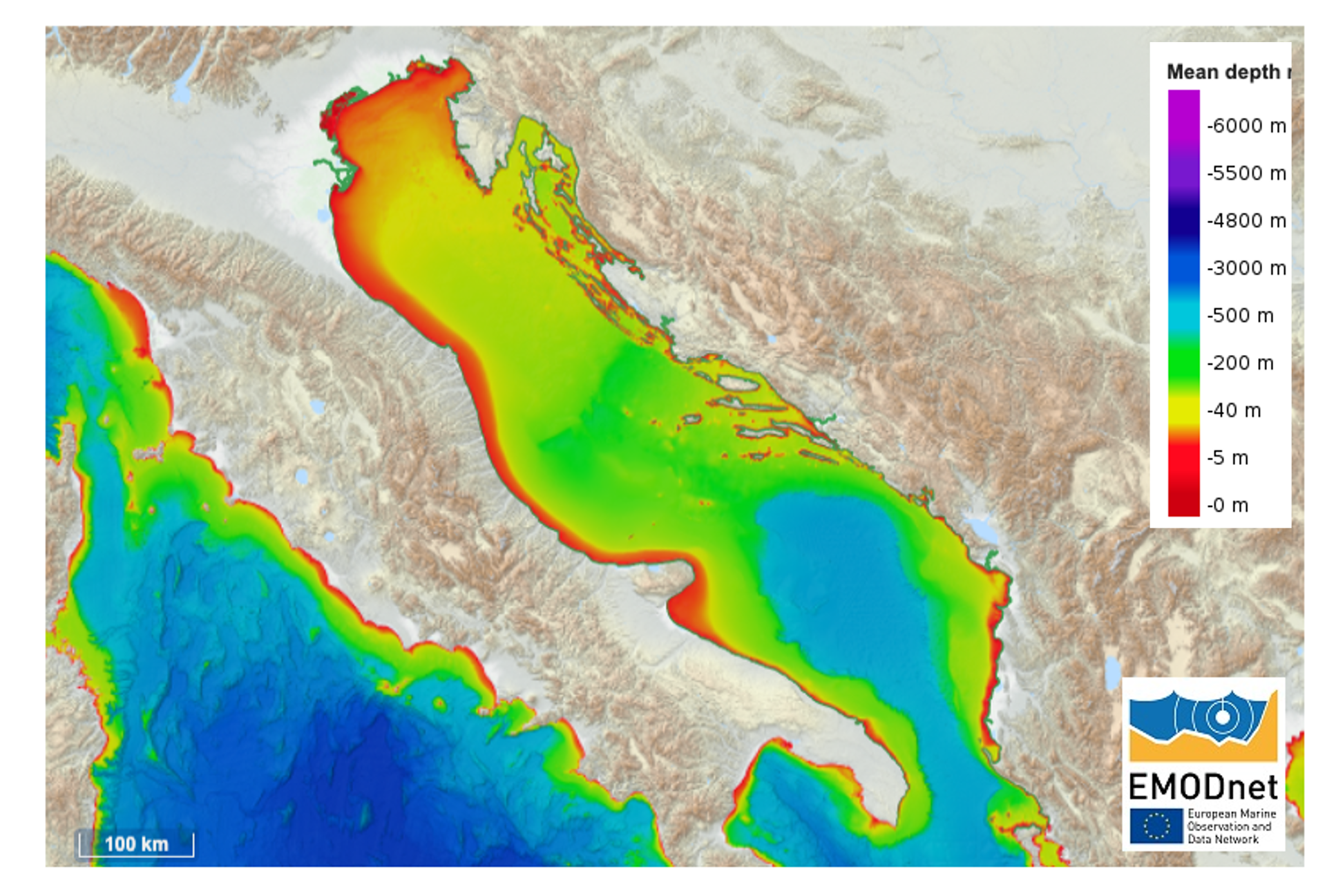}
    \caption{Adriatic Sea bathymetry from EMODnet Digital Bathymetry \cite{EMODnetDTM2024}. Colour shading indicates water depth (m)}
    \label{fig:Adriatic_sea}
\end{figure}

The study area considered in this work is the Adriatic Sea, which is a semi-enclosed basin in the northeastern Mediterranean. It divides the Italian and Balkan Peninsulas (see Figure \ref{fig:Adriatic_sea}). This region is one of the most dynamically complex sub-basins in the Mediterranean region. It is elongated along a northwest-southeast axis with a pronounced north-south bathymetric gradient. The northern shelf, extending from the Gulf of Venice to Ancona, is exceptionally shallow ($\approx 35$m) and is strongly influenced by wintertime atmospheric cooling and freshwater discharge from the Po River and other Apennine rivers. The central basin averages $\approx 140$m depth, with localised depressions at the Pomo (Jabuka) Pit exceeding 260~m. The southern Adriatic is dominated by the South Adriatic Pit (deeper than $1200$m) and communicates with the Ionian Sea through the Strait of Otranto (sill depth $\approx 800$m) \cite{Artegiani_1997,Bignami_2007}.

\subsection{Dataset}
High-resolution (HR) oceanographic fields were extracted from the Mediterranean Sea Physics Reanalysis (Med MFC) \cite{CMEMS_MedMFC_PHY}, which provides daily outputs on a 1/24$\degree$ ($\approx 4.5$km) horizontal grid with 141 unevenly spaced vertical levels. The dataset is generated by the NEMO hydrodynamic model combined with the OceanVAR variational assimilation scheme, which ingests vertical profiles of temperature and salinity together with satellite sea level anomaly observations.
Corresponding low-resolution (LR) data were extracted from the Global Ocean Physics Reanalysis (GLORYS12V1), a CMEMS global eddy-resolving product at 1/12$\degree$ ($\approx 8$km) horizontal resolution with 50 vertical levels. GLORYS12V1 assimilates along-track altimeter data, satellite sea surface temperature, sea ice concentration, and in situ temperature and salinity profiles using a reduced-order Kalman filter and 3D-VAR bias correction scheme. 
Four sea-surface variables are considered: temperature, salinity, zonal and meridional velocity components. For model training, HR reanalysis data and LR conditioning data spanning 1 January 2011 to 31 December 2020 were used, while data from 2009 were reserved for evaluation. This temporal split, in which the test year predates the training period, ensures strict independence from any temporal autocorrelation in the reanalysis products. 

\subsection{Data Preprocessing and Loading}
\label{sec:preprocessing}
A dedicated data loader was implemented in PyTorch to handle paired HR and LR datasets efficiently, interfacing directly with xarray datasets in NetCDF-4 format to enable on-demand, memory-efficient access during training.

Temporal consistency is enforced by selecting only timestamps common to both HR and LR datasets (Med MFC and GLORYS12V1). For each valid timestep, the four oceanographic variables are standardized through a two-step normalization procedure: (i) z-score normalization based on the combined mean and standard deviation computed across both HR and LR fields to preserve inter-resolution scale consistency, followed by (ii) min-max scaling to the range $[-1,1]$. Missing values are replaced with zeros to maintain uniform tensor dimensions and numerical stability.

A binary ocean mask, derived from the HR fields, is applied to restrict training and evaluation to valid ocean grid cells, excluding land areas from all metric computations. Seasonal variability is represented by a sinusoidal embedding of the month index, providing cyclic temporal context to the learning process.

\begin{figure}[ht]
    \centering
    \includegraphics[width=0.8\textwidth]{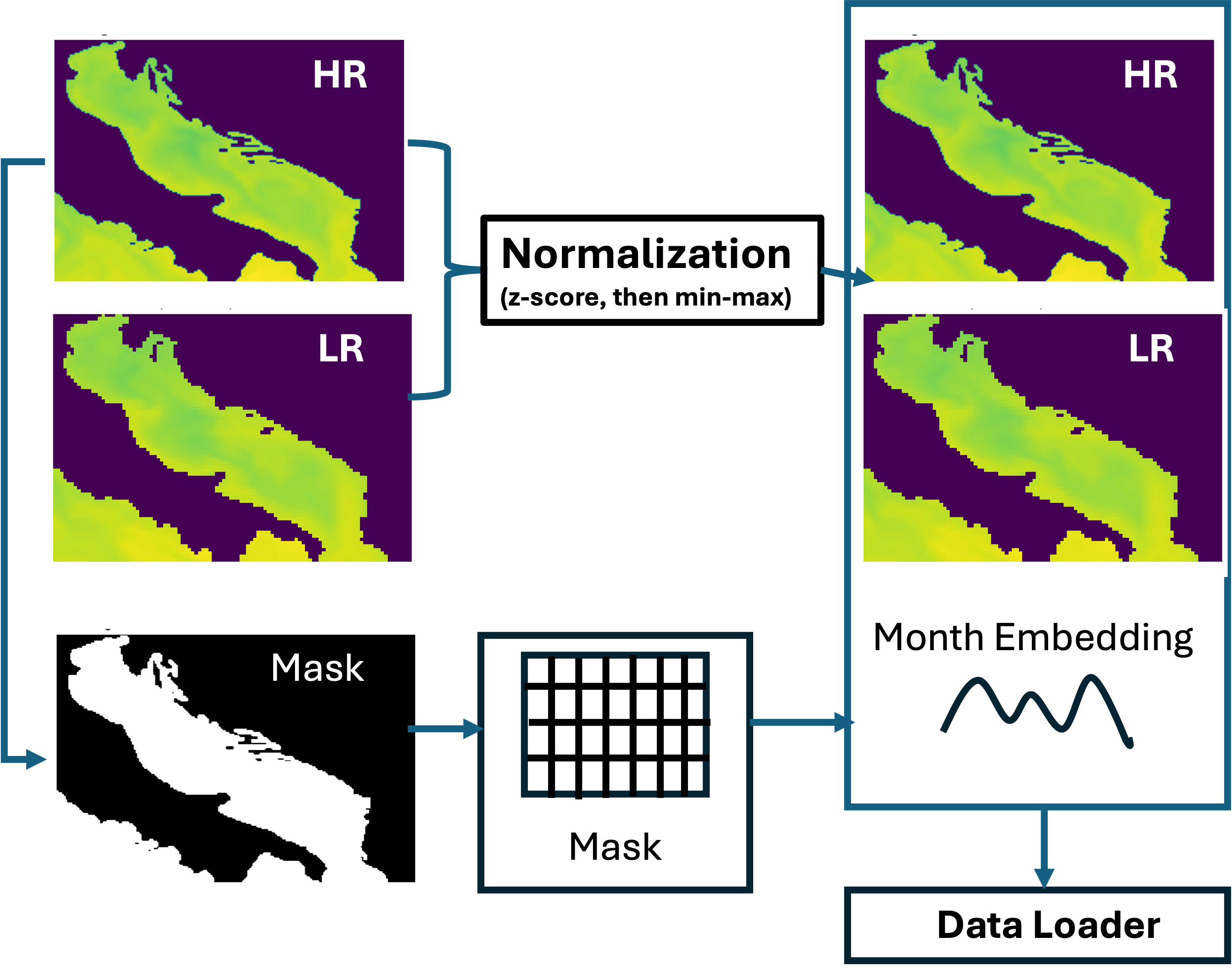}
    \caption{Schematic of the data loading and preprocessing pipeline used for paired high-resolution (HR) and low-resolution (LR) oceanographic datasets.}
    \label{fig:dataloader}
\end{figure}

The loader returns HR reanalysis tensors (ground truth), LR conditioning tensors, temporal embeddings, and ocean masks in a channel-first tensor shape, as summarized in Fig.~\ref{fig:dataloader}. Data are supplied through a multi-threaded data loader, allowing parallel I/O operations and efficient GPU utilization. This modular pipeline is designed to support training of conditional diffusion architectures for sub-grid-scale oceanographic reconstruction.

\subsection{Conditional U-Net for Spatiotemporal Super-Resolution}
\begin{figure}[ht]
    \centering
    \includegraphics[width=1.0 \textwidth]{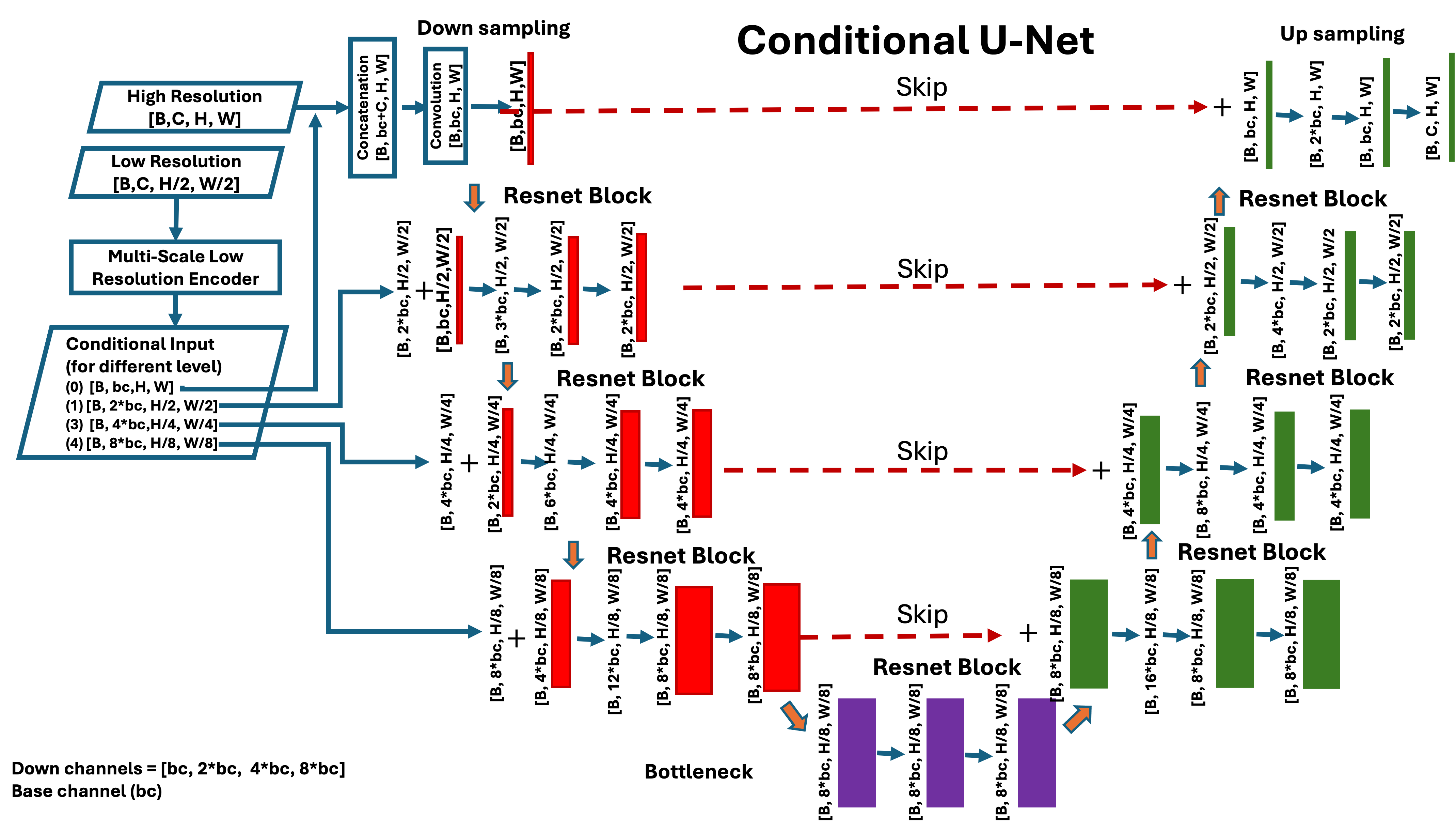}
    \caption{Schematic of the Conditional U-Net architecture used for spatiotemporal super-resolution of oceanographic parameters.}
    \label{fig:conditional_Unet}
\end{figure}

We adopt a conditional convolutional U-Net architecture as the noise-prediction network to reconstruct high-resolution oceanographic fields from corresponding low-resolution inputs while incorporating temporal and seasonal information with sinusoidal embeddings. The model follows a classical encoder-decoder design, enhanced with residual blocks, attention mechanisms, and multi-scale conditioning to capture both fine-scale structures and large-scale spatial coherence.
The encoder progressively downsamples the high-resolution input through a series of residual blocks with group normalization and SiLU activations. Skip connections preserve intermediate feature maps for later use in the decoder, enabling precise reconstruction of spatial details. At the bottleneck, a cross-attention module captures long-range dependencies, enabling the network to represent coherent large-scale structures such as basin-wide thermohaline gradients and mesoscale circulation patterns.
In the decoder, feature maps are successively upsampled and merged with corresponding encoder representations via skip connections. This combination of residual processing and feature fusion ensures that both global context and local structure are effectively retained during reconstruction.
A distinguishing design element is the multi-scale conditioning pathway: a dedicated low-resolution encoder processes the LR input and produces conditioning feature maps at each of the four block levels of the main U-Net. The resulting conditioning features are spatially aligned and concatenated with the main U-Net activations at each level, guiding the reconstruction with coarse-scale physical context.
Temporal information is incorporated through sinusoidal timestep embeddings and month embeddings, projected via lightweight multilayer perceptrons (MLPs), and injected into the network using Feature-wise Linear Modulation (FiLM). This allows the model to adapt its activations according to the temporal context, effectively capturing seasonal variability in oceanographic fields.
Formally, the diffusion process begins from pure Gaussian noise $\mathbf{x}_T\sim\mathcal{N}(\mathbf{0},\mathbf{I})$, and the conditional U-Net learns the mapping
\begin{equation}
    \hat{\boldsymbol{\epsilon}}_t = f_{\boldsymbol{\theta}}(\mathbf{x}_t, \mathbf{c}, t,\mathbf{d}) ,
\end{equation}

where $\mathbf{x}_t$  represents the high-resolution field at a given diffusion step, $\mathbf{c}$ denotes the low-resolution conditioning input, $t$ encodes the diffusion step, and $\mathbf{d}$ represents the month embedding. By iterating the learned reverse transitions over $T$ steps, OcDiffSR maps the initial Gaussian noise $\mathbf{x}_T$ back to the high-resolution data distribution, yielding the reconstructed field $\hat{\mathbf{x}}_0$.
Overall, the proposed conditional U-Net (presented in Fig. \ref{fig:conditional_Unet}) provides a framework for spatiotemporal super-resolution of oceanographic parameters, linking low-resolution conditional inputs with high-resolution reconstructions that retain both local detail and large-scale coherence.

\subsection{Evaluation Metrics}
The performance of the OcDiffSR model has been assessed using seven statistical image-based evaluation metrics between the model output and the reanalysis data which serves as ground truth. These metrics quantify the accuracy, reliability, and structural similarity of the generated oceanographic variables, namely the sea surface salinity $(\mathbf{s})$, sea surface temperature $(\boldsymbol{\vartheta})$, and velocity components $(\mathbf{u},\mathbf{v})$. The metrics are described below in detail.

\subsubsection{Root Mean Square Error (RMSE)}
The RMSE measures the mean squared deviation between the output of the model and reanalysis \cite{murphy1987general, wang2009mean}.
\begin{equation}
    \text{RMSE} = \sqrt{\frac{1}{N} \sum_{i=1}^N (G_i -R_i)^2},
\end{equation}

where $G_i$ and $R_i$ represent the generated and reanalysis data points, respectively, and $N$ is the total number of data points (pixels).  Lower values indicate better reconstruction accuracy, with RMSE~$= 0$ denoting perfect agreement.

\subsubsection{Mean Absolute Error (MAE)}
The MAE measures the average absolute difference between generated and reanalysis data. 
\begin{equation}
    \text{MAE} = \frac{1}{N} \sum_i^N|{G_i - R_i}|
\end{equation}
MAE is less sensitive to outliers than RMSE and provides a direct measure of the mean prediction error magnitude. A lower value of MAE indicates better performance.

\subsubsection{Pearson Correlation Coefficient (PCC)}
The PCC measures the linear correlation between the model output and reanalysis datasets \cite{wilks2011statistical}.
\begin{equation}
    \text{PCC} =\frac{\text{Cov}(G,R)}{\sigma_G \sigma_R} = \frac{\sum_{i=1}^N(G_i - \bar{G})(R_i-\bar{R})}{\sqrt{\sum_{i=1}^N{(G_i - \bar{G})^2}\sum_{i=1}^N{(R_i - \bar{R})^2}}},
\end{equation}

where $\bar{G}$ and $\bar{R}$ are the means of the model output and reanalysis data and $\sigma_G$ and $\sigma_R$ are their standard deviations.
The PCC values range from $-1$ to $+1$; values above $0.9$ indicate excellent spatial agreement, between $0.7$ and $0.9$ good agreement, and below $0.7$ poor agreement \cite{wilks2011statistical}.

\subsubsection{Coefficient of Determination}
The coefficient of determination quantifies how well the generated data explain the variance of the reanalysis data \cite{wilks2011statistical}.
\begin{equation}
    R^2 = 1- \frac{\sum_{i=1}^N (R_i-G_i)^2}{\sum_{i=1}^N(R_i -\bar{R})^2},
\end{equation}

$R^2 = 1$ indicates perfect agreement; values near zero indicate that the model performs no better than the climatological mean, and negative values indicate performance worse than the mean predictor. In oceanographic datasets, $R^2$ values greater than $0.9$ indicate excellent explanatory performance \cite{wilks2011statistical}.
\subsubsection{Structural Similarity Index Measure (SSIM)}
The SSIM evaluates the perceived similarity between two images based on structural information \cite{wang2009mean}.
\begin{equation}
    \text{SSIM}(G,R) = \frac{(2\mu_G\mu_R + C_1)(2\sigma_{GR} +C_2)}{(\mu_G^2 +\mu_R^2 +C_1)(\sigma_G^2+\sigma_R^2 +C_2)},
\end{equation}

where $\mu_G$, $\mu_R$ are the local mean, $\sigma_G^2$ and $\sigma_R^2$ are the variances, $\sigma_{GR}$ is the covariance, $C_1$ and $C_2$ are small constants to stabilize the division.
The SSIM values range from $0$ to $1$, with $1$ indicating perfect structural similarity. Typically, SSIM$> 0.9$ indicates excellent structural similarity, between $0.7-0.9$ good, and below $0.5$ is considered poor similarity.

\subsubsection{Peak Signal-to-Noise Ratio (PSNR)}
The PSNR measures the ratio between the maximum possible signal power and the noise introduced by reconstruction errors \cite{wang2004imagequality}\cite{gonzalez2008digital}.

\begin{equation}
    \text{PSNR} = 10 \log_{10}\left(\frac{\text{MAX}^2}{\text{MSE}}\right)
\end{equation}

where MSE is the mean squared error between the reconstructed and reference fields, and  MAX denotes the maximum possible value of the signal in the representation used for evaluation. In this work, all oceanographic fields undergo a two-step normalisation before model training and evaluation: first, z-score standardisation based on the combined mean and standard deviation computed across both HR and LR fields, followed by min-max scaling to the range $[-1, 1]$ (Section~\ref{sec:preprocessing}).  Consequently, the peak signal value is fixed at $\text{MAX} = 1$, corresponding to the upper bound of the normalised range, and this value is adopted uniformly across all variables and all PSNR computations reported in this study. For oceanographic data, PSNR~${> 30}$~dB indicates high fidelity, between $20$~dB and $30$~dB moderate, and less than $20$~dB poor. These generic fidelity bands are conventionally calibrated to data referenced against the full non-negative dynamic range (e.g.\ $[0,1]$); under the $[-1,1]$ convention with $\text{MAX}=1$ adopted here, an equivalent reconstruction error yields PSNR values approximately $6$~dB lower than under that convention, so the PSNR values reported in this study are, if anything, conservative relative to these literature benchmarks.

\subsubsection{Normalized Root Mean Square Error (NRMSE)}
The NRMSE normalizes RMSE by the data range, providing a scale-independent measure of performance \cite{wilks2011statistical}.
\begin{equation}
    \text{NRMSE}=\frac{\text{RMSE}}{(R_{\text{max}} -R_{\text{min}} )}
\end{equation}

NRMSE values closer to $0$ indicate better agreement, while values greater than $0.2$ generally represent poor correspondence between the generated and reference data.

\subsection{Bilinear interpolation}
Bilinear interpolation is a non-learnable rescaling technique widely used in image processing and numerical modelling to increase the spatial resolution of gridded fields \cite{gonzalez2008digital}. The method estimates the intensity value of a new pixel as the weighted average of the four nearest-neighbor pixels in the input grid. While it preserves spatial continuity, bilinear interpolation cannot recover sub-grid variability or nonlinear structures and is used here solely as a non-parametric baseline.

Mathematically, for a 2D input $I(x,y)$, the interpolated pixel value $I'(x',y')$ at fractional coordinate $(x',y')$ within a normalized unit square, where $(x', y') \in [0,1]$, is defined as:

\begin{equation}
    I'(x',y' )=\sum_{i=0}^1\sum_{j=0}^1 w_{ij} I(x_i,y_j)
\end{equation}

where $(x_i,y_i)$ represent the coordinates of the four nearest pixels surrounding the point $(x',y')$, and $w_{ij}$ are the interpolation weights given by:
\begin{equation}
    w_{ij}=(1-|x'-x_i|)(1-|y'-y_j|)
\end{equation}

These weights are valid under the assumption that the local coordinates $(x',y')$ and $(x_i,y_j)$ are normalized within a unit square cell, ensuring that the interpolated value transitions smoothly between the known pixel intensities. Compared to nearest-neighbor interpolation, bilinear interpolation provides smoother and more visually coherent outputs by accounting for both horizontal and vertical variations from surrounding pixels.

\begin{figure}[ht]
    \centering
    \includegraphics[width=0.50 \textwidth]{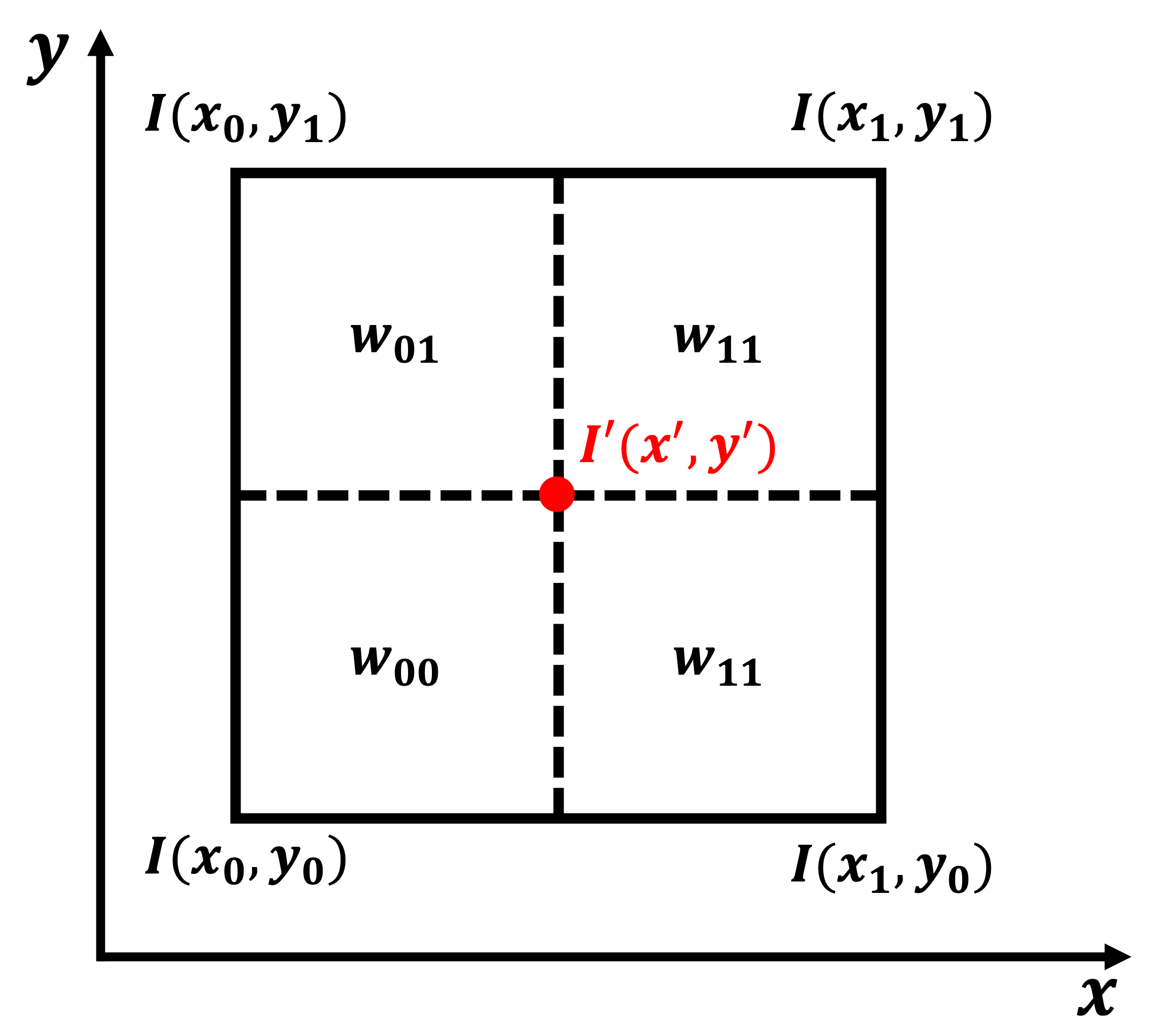}
    \caption{Schematic representation of bilinear interpolation on a normalized unit square.}
    \label{fig:bilinear_interpolation}
\end{figure}
Figure \ref{fig:bilinear_interpolation} illustrates the computation of the interpolated value $I'(x',y')$ within a unit square defined by the corner coordinates $(x_0,y_0)$, $(x_0,y_1)$, $(x_1,y_0)$, and $(x_1,y_1)$. The point $(x',y')$ lies inside this square, and its interpolated intensity is determined as a weighted average of the four neighboring pixel values, with weights $w_{00}$, $w_{01}$, $w_{10}$, and $w_{11}$ corresponding to the four corners. The dashed lines indicate the relative distances from $(x', y')$ to each corner in the $x$- and $y$-directions, which determine the interpolation weights $w_{ij}$.
\subsection{Residual corrective diffusion modeling (CorrDiff)}

The Residual Corrective Diffusion (CorrDiff) model~\cite{mardani2025} is a two-stage generative downscaling framework that combines a U-Net-based regression model with a conditional diffusion model to represent the distribution of high-resolution atmospheric fields conditioned on coarse-resolution inputs. The method is specifically designed to address multiscale challenges inherent in atmospheric downscaling, including the representation of fine-scale variability and extreme weather phenomena. CorrDiff decomposes the target field into a deterministic conditional mean and a stochastic residual. In the first stage, a convolutional UNet is trained using a regression objective to approximate the conditional expectation of the high-resolution state, effectively capturing large-scale and quasi-deterministic physical relationships. In the second stage, a conditional diffusion model is applied to the residual between the target field and the predicted mean, enabling stochastic generation of unresolved small-scale structures and physical processes. By modeling the residual rather than the full field, CorrDiff reduces the variance and complexity of the generative task, yielding improved training stability, more efficient sampling, and enhanced representation of small-scale variability. Final high-resolution realizations are obtained by combining the regressed mean with samples drawn from the residual diffusion model, yielding probabilistic and physically consistent downscaled outputs. 
To ensure a fair and rigorous comparison with OcDiffSR, CorrDiff was retrained from scratch on the same oceanographic dataset used in this study, rather than relying on its original atmospheric pre-trained weights. Specifically, the atmospheric data loader in the original CorrDiff implementation was replaced with the same oceanographic data loader used in OcDiffSR, which interfaces directly with paired GLORYS12V1 low-resolution and Med MFC high-resolution NetCDF-4 reanalysis fields over the Adriatic Sea. This substitution ensures that both models ingest identical training samples, the same four oceanographic variables (sea surface temperature, salinity, zonal and meridional velocity components), the same temporal split (training on 2011--2020, evaluation on 2009), the same two-step normalisation procedure (z-score followed by min-max scaling to $[-1, 1]$), and the same binary ocean mask excluding land grid cells from all computations. The CorrDiff model architecture and training objective were otherwise kept identical to the original formulation \cite{mardani2025}, with no modifications to the diffusion backbone, regression stage, or residual decomposition strategy. This retraining procedure guarantees that any observed differences in reconstruction performance between CorrDiff and OcDiffSR are attributable solely to architectural and methodological design choices, rather than to differences in training data, preprocessing, or domain coverage.

\subsection{Model Architecture and Training Configuration}
OcDiffSR model is implemented using a Conditional U-Net architecture with an explicit channel configuration of [64, 128, 256, 512], enabling a progressive increase in feature dimensionality across resolution levels, with a maximum of 512 channels. Each resolution level consists of two residual blocks, and conditional super-resolution is achieved by incorporating low-resolution input fields as conditioning channels. Temporal information is encoded using 128-dimensional diffusion time embeddings together with 2-dimensional date embeddings, resulting in a total of 49.9 million trainable parameters. Training was performed in a distributed data-parallel configuration across four NVIDIA A100 GPUs using the NCCL communication backend on the Leonardo supercomputer at CINECA, Italy.

The CorrDiff model employs a diffusion U-Net architecture with a base channel width of 128 and channel scaling defined by channel multipliers [1, 2, 2], yielding a maximum of 256 channels. Temporal conditioning is provided through diffusion time embeddings, and self-attention is applied at a spatial resolution of 16×16 to capture long-range spatial dependencies. The CorrDiff architecture contains 41.99 million trainable parameters.
The parameter counts of OcDiffSR (49.9~million) and CorrDiff (41.99~million) are comparable, ensuring a fair model-capacity comparison. CorrDiff's other hyperparameters, including the learning rate, training schedule, noise schedule, and the relative allocation between the regression and diffusion stages, were kept at their originally published values \cite{mardani2025}; the parameter-count matching described above was the only tuning performed for this comparison, and no additional oceanography-specific hyperparameter search was carried out for CorrDiff.

Unless otherwise stated, all high-resolution reconstructions reported in this study, including the quantitative results in Table~\ref{tab:quantitative_comparison} and the daily/monthly evaluation metrics, are generated by running the full $T = 1000$-step reverse diffusion process at inference time, matching the number of forward-process steps used during training. Each reported reconstruction corresponds to a single sampled realisation of the reverse process rather than an average over multiple independent samples; characterising the spread across repeated stochastic samples for a fixed conditioning input is left to future work.

\section{Results and Discussion}
\begin{figure}[htbp]
  \centering
  \includegraphics[width=\textwidth]{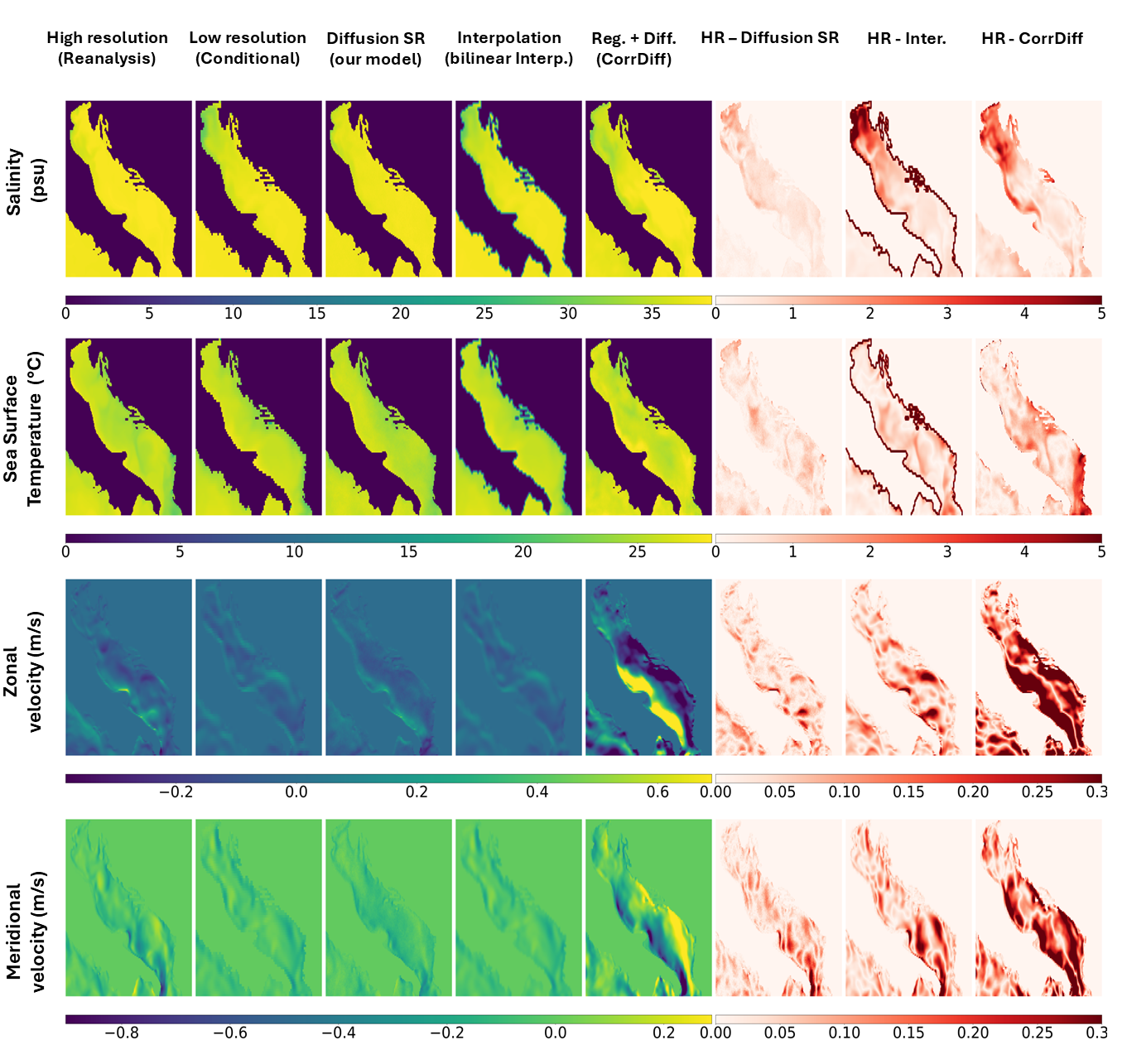}
  \caption{Qualitative comparison of super-resolution results for oceanographic variables on \textbf{7 August 2009} — the best-performance day of OcDiffSR over the 2009 test year (mean NRMSE averaged across all four variables $= 0.0243$) — over the Adriatic Sea domain. From left to right, each row shows: high-resolution reanalysis (ground truth), low-resolution conditional input, OcDiffSR output (our model), bilinear interpolation, and regression-plus-diffusion (CorrDiff). The last three columns report absolute error maps with respect to the reanalysis for the OcDiffSR, bilinear interpolation and CorrDiff models, respectively. Rows correspond to (top to bottom) salinity (psu), sea surface temperature ($^{\circ}$C), zonal velocity ($\mathrm{m\,s^{-1}}$), and meridional velocity ($\mathrm{m\,s^{-1}}$). Colorbars beneath each row indicate the physical value ranges (left) and the corresponding error magnitudes (right).}
  \label{fig:best_day}
\end{figure}

\begin{figure}[htbp]
  \centering
  \includegraphics[width=\textwidth]{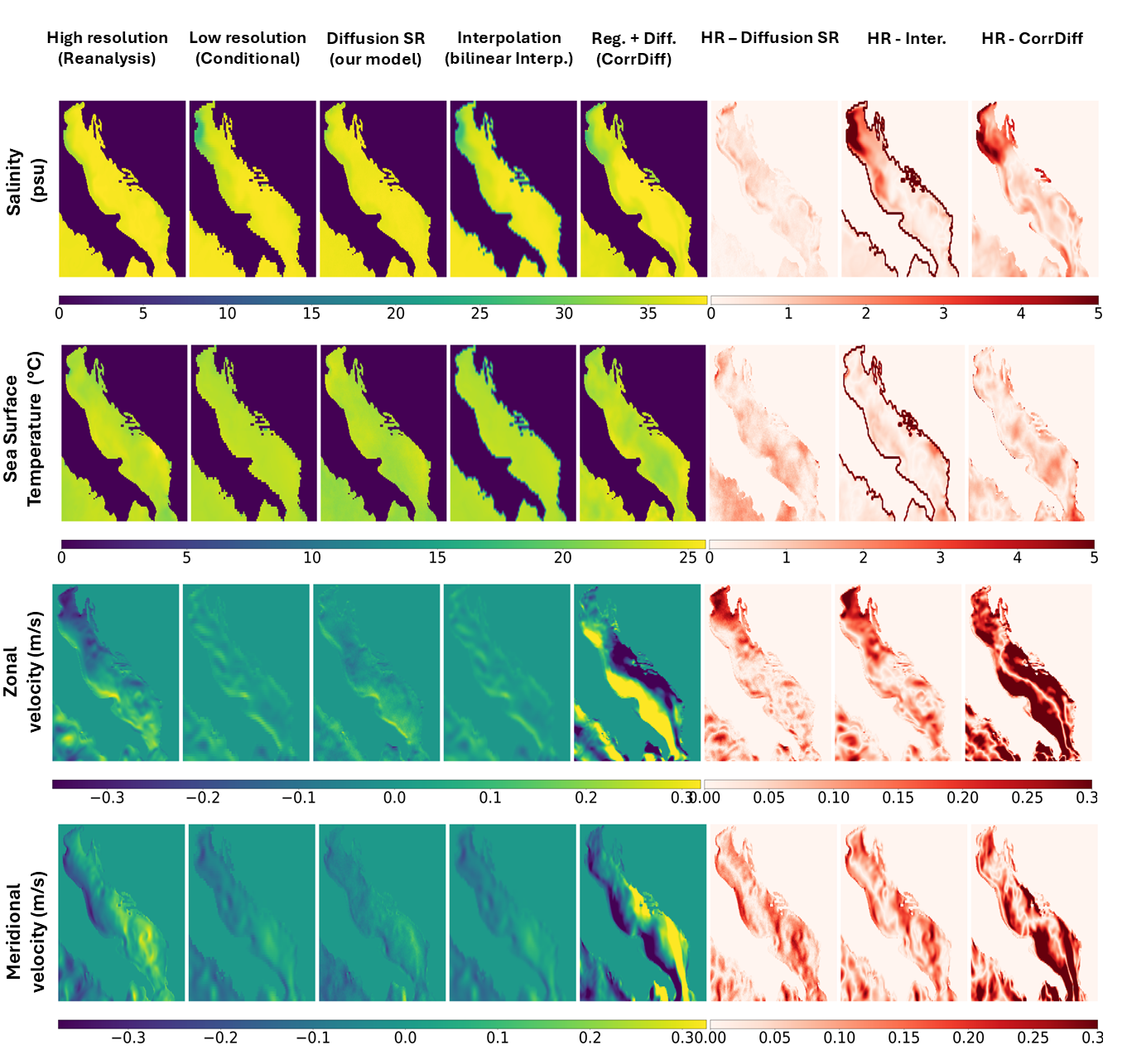}
  \caption{Qualitative comparison of super-resolution results for oceanographic variables on \textbf{27 May 2009} — the worst-performance day of OcDiffSR over the 2009 test year (mean NRMSE averaged across all four variables $= 0.0516$) — over the Adriatic Sea domain. From left to right, each row shows: high-resolution reanalysis (ground truth), low-resolution conditional input, OcDiffSR output (our model), bilinear interpolation, and regression-plus-diffusion (CorrDiff). The last three columns report absolute error maps with respect to the reanalysis for the OcDiffSR, bilinear interpolation, and CorrDiff models, respectively. Rows correspond to (top to bottom) salinity (psu), sea surface temperature ($^{\circ}$C), zonal velocity ($\mathrm{m\,s^{-1}}$), and meridional velocity ($\mathrm{m\,s^{-1}}$). Colorbars beneath each row indicate the physical value ranges (left) and the corresponding error magnitudes (right).}
  \label{fig:worst_day}
\end{figure}

The OcDiffSR model was trained for 2000 epochs on daily oceanographic fields spanning 1 January 2011 to 31 December 2020 (10 years), achieving a final training loss of $5 \times 10^{-4}$. Model evaluation was then performed on the independent test year 2009. To characterise the full range of reconstruction quality exhibited by OcDiffSR across the test year, qualitative results are presented for two representative days selected based on the mean Normalised Root Mean Square Error (NRMSE) averaged across all four oceanographic variables (sea surface salinity, sea surface temperature, zonal velocity, and meridional velocity): the \textbf{best-performance day}, 7 August 2009 (mean NRMSE $= 0.0243$; Figure~\ref{fig:best_day}), and the \textbf{worst-performance day}, 27 May 2009 (mean NRMSE $= 0.0516$; Figure~\ref{fig:worst_day}). These two cases bracket the reconstruction quality of OcDiffSR over the entire 365-day evaluation period and provide a physically interpretable account of the conditions under which the model excels and where it faces the greatest challenge. For both days, the high-resolution Mediterranean Sea Physics Reanalysis (Med~MFC) dataset is treated as the reference ground truth, while the low-resolution Global Ocean Physics Reanalysis (GLORYS12V1) dataset serves as the conditional input to the super-resolution models. The evaluation is conducted for four oceanographic variables (sea surface temperature ($\vartheta$), sea surface salinity ($s$), zonal velocity ($\mathbf{u}$), and meridional velocity ($\mathbf{v}$)) over the Adriatic Sea. Results are benchmarked against bilinear interpolation as a non-parametric baseline and the state-of-the-art residual diffusion model CorrDiff\cite{mardani2025} as a reference method.

Figure~\ref{fig:best_day} presents the qualitative comparison for 7 August 2009, the day on which OcDiffSR achieves its lowest single-day mean NRMSE across all four variables (mean NRMSE $= 0.0243$). This date falls in the peak summer season, when the Adriatic Sea is characterised by strong thermal stratification, well-defined mesoscale structures, and relatively stable large-scale circulation patterns — conditions that are well-represented in the ten-year training climatology and are therefore most amenable to accurate reconstruction.

Sea surface salinity (shown in the first row of Figure~\ref{fig:best_day}) is particularly challenging to reconstruct owing to its highly localised sources and sinks (evaporation, precipitation, and runoff). The dominant salinity patterns are faithfully reproduced by OcDiffSR, as confirmed by the error map (HR $-$ Diffusion SR), which shows spatially compact, low-magnitude residuals across the basin. Notably, errors remain relatively small near the coastline, where strong gradients and land-sea masking effects usually make super-resolution the most difficult. In contrast, bilinear interpolation and CorrDiff both exhibit structurally larger and more spatially heterogeneous errors on this date.

For sea surface temperature, the low-resolution input exhibits a broad, smooth thermal gradient, whereas the reanalysis data reveal a high-resolution thermal structure with sharp frontal features and mesoscale gradients. The OcDiffSR model effectively reconstructs these fine-scale SST patterns, producing realistic spatial distributions with minimal absolute error across the open basin. A slight cold bias is nonetheless evident in the immediate vicinity of the coastline. This artefact is a direct consequence of the land-sea masking strategy adopted during preprocessing: missing values at land grid cells are replaced with zero prior to normalisation (Section~\ref{sec:preprocessing}), introducing a sharp discontinuity at the land-sea boundary that the model learns to partially reproduce. Importantly, this bias is spatially confined to the coastal fringe and does not affect open-ocean reconstructions; furthermore, since the binary ocean mask explicitly excludes all land cells from metric computations (Section~\ref{sec:preprocessing}), the reported quantitative scores are not inflated by this boundary effect. Addressing this limitation through physically motivated coastal fill strategies, such as nearest-ocean-neighbour extrapolation or climatological boundary padding, is identified as a priority for future refinement of the preprocessing pipeline. Nevertheless, the model maintains excellent agreement with the high-resolution Med~MFC reanalysis data across all ocean grid cells.

The third row represents zonal velocity. OcDiffSR successfully reconstructs coherent flow structures, including localised jets absent from the coarse GLORYS12V1 input but present in the Med~MFC reanalysis. The zonal velocity field generated from OcDiffSR recovers these features, displaying spatially coherent flows. The difference map (HR - OcDiffSR) shows low-magnitude residuals, suggesting minor phase or amplitude errors rather than missing flow.
Similarly, in the case of meridional velocity (fourth row), the conditional low-resolution data are smoothed, with minimal representation of north-south currents. The OcDiffSR model reconstructs the meridional velocity in good agreement with the high-resolution Med MFC reanalysis dataset, demonstrating strong generalization for directionally anisotropic flow. The error map shows residuals concentrated along dynamic boundaries, suggesting that OcDiffSR tends to underestimate small-scale vorticity and exhibits phase errors consistent with those observed for the zonal component. 

Overall, OcDiffSR successfully reproduces high-resolution oceanographic structures across all four variables on 7 August 2009. The model reconstructs coherent thermal and haline patterns with exceptional sharpness and physical realism in the case of scalar variables (sea surface temperature and salinity). In the case of zonal and meridional velocities, OcDiffSR generates coherent data similar to that of the high-resolution reanalysis dataset. It shows dynamically consistent circulation patterns that align with the mesoscale features observed in the reanalysis. The difference maps (HR - OcDiffSR) confirm that the residual errors are small, spatially structured and are mainly due to feature displacement (phase errors) rather than missing information.

Figure \ref{fig:best_day} also presents a qualitative comparison of different super-resolution models, taking bilinear interpolation as a baseline and CorrDiff \cite{mardani2025} as a reference method for the OcDiffSR model. As expected, bilinear interpolation produces spatially smooth fields but fails to recover fine-scale structures and sharp gradients. This behavior is evident across all variables, particularly in coastal regions and frontal zones, where interpolation excessively smooths spatial variability and leads to large, structured errors relative to the high-resolution reanalysis.

The OcDiffSR model shows a clear improvement over bilinear interpolation for all variables. For salinity and sea surface temperature, OcDiffSR better reconstructs coastal gradients and mesoscale features that are largely absent in the interpolated fields. The residual map (HR - OcDiffSR) shows substantially lower error magnitudes and improved spatial coherence when compared to the baseline interpolation model, indicating better agreement with the reanalysis.

The OcDiffSR model also demonstrates similar or superior performance to CorrDiff,  particularly in terms of spatial consistency and error distribution. Although the CorrDiff model can recover some fine-scale variability better than that of interpolation, it frequently introduces amplified errors and localized artifacts, particularly in areas with strong gradients and close proximity to coastlines. The error magnitudes in the CorrDiff absolute error maps are systematically higher and more spatially heterogeneous than those of OcDiffSR.

The limitations of bilinear interpolation are more noticeable for the dynamical variables (meridional and zonal velocities), as it is unable to reconstruct small-scale variability and coherent current structures. The CorrDiff model also struggles to reproduce coherent filamentary structures, resulting in higher-magnitude and more fragmented residuals relative to the high-resolution Med MFC reanalysis. In contrast, OcDiffSR more effectively recovers coherent velocity features and yields smoother, physically consistent fields, resulting in the lowest error levels among the three methods.

Overall, relative to bilinear interpolation and CorrDiff, the OcDiffSR model produces fields that are visually closer to the high-resolution reanalysis, with sharper structures, improved representation of mesoscale dynamics, and systematically reduced errors. 

Figure~\ref{fig:worst_day} presents the qualitative comparison for 27 May 2009, the day on which OcDiffSR achieves its highest single-day mean NRMSE across all four variables (mean NRMSE $= 0.0516$). This date falls in the late-spring transitional period, when the Adriatic Sea undergoes the breakdown of winter vertical mixing and the onset of summer stratification, producing transient mesoscale structures and dynamically complex circulation that are intrinsically more challenging to reconstruct than the stable summer conditions observed on the best-performance day.

Sea surface salinity (first row of Figure~\ref{fig:worst_day}) remains broadly well-reproduced by OcDiffSR, which still captures the dominant large-scale salinity distribution; however, the error map reveals modestly larger residuals compared to 7 August, particularly along the western coastal shelf where freshwater runoff from Apennine rivers interacts with offshore water masses during this transitional season. Despite this degradation, OcDiffSR continues to exhibit lower and more spatially compact errors than both bilinear interpolation and CorrDiff on this date.

For sea surface temperature (second row of Figure~\ref{fig:worst_day}), the transitional thermal structure of late May characterised by partially formed fronts and heterogeneous surface warming poses a greater reconstruction
challenge than the fully stratified summer conditions of August. OcDiffSR recovers the broad thermal gradient and major mesoscale features, but fine-scale frontal sharpness is somewhat reduced relative to the best-performance day, as reflected in modestly elevated error magnitudes visible in the difference map. A slight cold bias near the coastal fringe remains present, arising from the same land-sea masking artefact described in the context of 7 August. Nevertheless, OcDiffSR maintains lower SST errors than both bilinear interpolation and CorrDiff; the latter introduces amplified artifacts near frontal zones, leading to systematically larger and more spatially heterogeneous error magnitudes.

The third row of Figure~\ref{fig:worst_day} represents the zonal velocity. The late-spring circulation in the Adriatic is characterised by enhanced mesoscale eddy activity and transitional current reversals driven by changing wind forcing and buoyancy fluxes. These transient, high-variability flow structures are the most challenging features for any super-resolution model to reconstruct, as they involve fine-scale spatial patterns with limited low-resolution precursors in the GLORYS12V1 conditioning input. The zonal velocity error map shows increased residual magnitudes relative to August, concentrated along dynamically active boundaries and mesoscale eddy margins.
Nonetheless, OcDiffSR continues to outperform both baselines: bilinear interpolation fails to recover any coherent current structures, while CorrDiff produces the highest error magnitudes and most spatially fragmented residuals among the three methods. In the case of meridional velocity (fourth row), the increased spatial intermittency of north-south flow during the transitional season leads to larger phase errors in the OcDiffSR output, with residuals
visible along the western boundary current and within mesoscale eddy margins. The error map shows residuals concentrated along dynamic boundaries, consistent with the phase error behaviour observed for the zonal component. Nevertheless, the OcDiffSR error magnitudes remain substantially lower than those of CorrDiff, which exhibits the largest and most spatially heterogeneous residuals among all three methods on this date.

Overall, OcDiffSR successfully reproduces high-resolution oceanographic structures across all four variables on 27 May 2009, even under the most dynamically challenging conditions encountered in the test year. The model maintains physically coherent fields and lower errors than both bilinear interpolation and CorrDiff across all four variables, confirming the robustness of the model across the full range of Adriatic Sea dynamical regimes.

The contrast between the 7 August and 27 May results reveals that OcDiffSR performance is modulated primarily by the dynamical complexity of the ocean state rather than by scalar field amplitude. Summer conditions, characterised by stable stratification and well-organised mesoscale structures, yield the most accurate reconstructions (mean NRMSE $= 0.0243$). Late-spring transitional periods, when the ocean simultaneously supports residual winter mixing, emergent stratification, and energetic mesoscale activity, present the greatest reconstruction challenge (mean NRMSE $= 0.0516$). The ratio between the worst- and best-day mean NRMSE is approximately 2.1, indicating that model degradation is bounded and physically interpretable. Crucially, even on the worst-performance day, OcDiffSR produces fields that are visually closer to the high-resolution reanalysis than both bilinear interpolation and CorrDiff, with sharper structures, improved representation of mesoscale dynamics, and systematically reduced errors.

To complement the qualitative analysis, a comprehensive statistical evaluation was conducted across all 365~days of the test year. Performance metrics were computed for the outputs generated by OcDiffSR, the interpolation baseline, and the CorrDiff model and were compared against the high-resolution Med-MFC reanalysis dataset for the whole year (inference dataset year 2009). The evaluation was performed for key oceanographic variables, including sea surface salinity, sea surface temperature, and zonal and meridional velocities. The performance was assessed using standard image-based metrics: MAE, NRMSE, PCC, PSNR, $R^2$, RMSE, and SSIM. These metrics provide complementary insights: RMSE and MAE quantify absolute errors; NRMSE accounts for scale-independent error; PCC and $R^2$ indicate linear correlation and variance explained; SSIM evaluates spatial-structural similarity; and PSNR measures overall reconstruction fidelity. The results are summarized in Table \ref{tab:quantitative_comparison}.

\begin{table*}[htbp]
\centering
\caption{Quantitative comparison of evaluation metrics across oceanographic variables for OcDiffSR, interpolation baseline, and CorrDiff model.}
\label{tab:quantitative_comparison}
\renewcommand{\arraystretch}{1.2}
\begin{tabular}{lccc}
\hline
\textbf{Metric} & \textbf{OcDiffSR} & \textbf{Interpolation} & \textbf{CorrDiff} \\
\hline
\multicolumn{4}{l}{\textbf{Sea surface salinity ($\mathbf{s}$)}} \\
\hline
MAE   & 0.1718 & 1.3078 & 0.3752 \\
NRMSE & 0.0089 & 0.1138 & 0.0228 \\
PCC   & 0.9999 & 0.9711 & 0.9990 \\
PSNR  & 41.0223 & 18.8727 & 32.8809 \\
$R^2$ & 0.9996 & 0.9429 & 0.9976 \\
RMSE  & 0.3455 & 4.3972 & 0.8832 \\
SSIM  & 0.9946 & 0.8840 & 0.9827 \\
\hline
\multicolumn{4}{l}{\textbf{Sea surface temperature ($\boldsymbol{\vartheta}$)}} \\
\hline
MAE   & 0.2420 & 0.6862 & 0.3944 \\
NRMSE & 0.0234 & 0.1023 & 0.0435 \\
PCC   & 0.9991 & 0.9724 & 0.9960 \\
PSNR  & 32.8014 & 19.8074 & 28.2435 \\
$R^2$ & 0.9970 & 0.9447 & 0.9873 \\
RMSE  & 0.4769 & 2.1473 & 0.8134 \\
SSIM  & 0.9639 & 0.8700 & 0.9373 \\
\hline
\multicolumn{4}{l}{\textbf{Zonal velocity component ($\mathbf{u}$)}} \\
\hline
MAE   & 0.0197 & 0.0256 & 0.0883 \\
NRMSE & 0.0514 & 0.0659 & 0.2506 \\
PCC   & 0.6456 & 0.3205 & 0.5788 \\
PSNR  & 25.9449 & 23.7468 & 12.6387 \\
$R^2$ & 0.3994 & 0.0226 & -14.3573 \\
RMSE  & 0.0439 & 0.0568 & 0.2083 \\
SSIM  & 0.6962 & 0.6259 & 0.5860 \\
\hline
\multicolumn{4}{l}{\textbf{Meridional velocity component ($\mathbf{v}$)}} \\
\hline
MAE   & 0.0208 & 0.0275 & 0.0524 \\
NRMSE & 0.0553 & 0.0700 & 0.1471 \\
PCC   & 0.6363 & 0.3456 & 0.5458 \\
PSNR  & 25.2720 & 23.2065 & 17.0186 \\
$R^2$ & 0.3743 & 0.01375 & -3.6188 \\
RMSE  & 0.0463 & 0.0590 & 0.1205 \\
SSIM  & 0.6946 & 0.6176 & 0.6166 \\
\hline
\end{tabular}
\end{table*}

For sea-surface salinity (Table~\ref{tab:quantitative_comparison}), OcDiffSR achieves the best performance among all three methods.  The NRMSE$=0.0089$ further confirms that errors are negligible relative to the range of salinity values. The Pearson correlation coefficients obtained by OcDiffSR are $0.9999$ and $R^2=0.9996$, 
indicating near-perfect linear correlation and explained variance, showing that the model reproduces the overall spatial pattern almost exactly. SSIM$=0.995$ confirms that spatial structures such as fronts and salinity gradients are preserved. The PSNR of 41.0 dB, the highest among the three models, indicates high-fidelity reconstruction with low noise relative to the reanalysis field.

CorrDiff outperforms bilinear interpolation but does not reach the performance of OcDiffSR on any of the reported metrics. The RMSE and MAE of  the CorrDiff model are $0.883$psu and $0.375$psu, respectively, and its PCC$=0.9990$ and SSIM$=0.983$ indicate good correlation and structural preservation. However, these values are slightly lower than those of OcDiffSR, suggesting that some fine-scale features are smoothed. Bilinear interpolation performs the worst, with RMSE$=4.397$psu, MAE$=1.308$psu, PCC$=0.9711$ and SSIM$=0.884$, reflecting strong smoothing and loss of small-scale variability. Overall, OcDiffSR preserves both accuracy and spatial coherence better than CorrDiff and interpolation.

For sea surface temperature, OcDiffSR achieves RMSE$=0.477\degree $C and MAE$=0.242\degree$C, showing minimal absolute error. The low NRMSE$=0.023$ confirms scale-independent accuracy. PCC$=0.9991$ and $R^2=0.9970$ indicate almost perfect correlation and variance capture, while SSIM$=0.964$ demonstrates excellent preservation of thermal fronts and mesoscale gradients, signifying strong structural consistency. The PSNR of $32.8$dB further confirms high-fidelity reconstruction, exceeding the $30$dB threshold.  Bilinear interpolation produced RMSE$=2.147\degree$C, MAE$=0.686\degree$C, and SSIM$=0.870$. The method exhibits oversmoothing of sharp fronts, resulting in slightly reduced spatial coherence and correlation with the reference field (PCC$=0.9724$).
CorrDiff performs better than interpolation with RMSE$=0.813\degree$C, MAE$=0.394\degree$C and SSIM$=0.937$, thus capturing overall patterns and some gradients; however, finer thermal structures are less accurately reproduced. OcDiffSR, therefore, provides superior accuracy, correlation, and structural fidelity.

The zonal $(\mathbf{u})$ and meridional $(\mathbf{v})$ velocity components, inherently more dynamic and less spatially coherent than scalar variables \cite{guillaumin2021stochastic}, display relatively lower metric values. The OcDiffSR model yields RMSE$=0.043$~m\,s$^{-1}$ and MAE$=0.019$~m\,s$^{-1}$ values for the zonal velocity $(\mathbf{u})$. PCC$=0.645$ and $R^2=0.399$ obtained from OcDiffSR indicate only a moderate and poor linear correlation with the reanalysis data. SSIM$=0.696$ points to fair structural similarity, suggesting that while the overall flow direction and major current features are preserved, finer-scale velocity fluctuations are more challenging for the model to reconstruct. The PSNR ($25.94$dB) remains above the moderate-poor fidelity threshold, and the NRMSE$=0.051$ implies slightly higher relative error. These results reflect the intrinsic difficulty of resolving vector fields characterized by high spatial variability and sensitivity to noise \cite{cheng2023machine}. Bilinear interpolation produces slightly higher absolute errors (RMSE$= 0.057$~m\,s$^{-1}$, MAE$= 0.026$~m\,s$^{-1}$) and fails to capture coherent flow patterns, as reflected by PCC$= 0.321$ and SSIM$= 0.626$. The CorrDiff model performs worst among the three methods for zonal velocity, exhibiting the highest error magnitudes, i.e., RMSE$=0.2083$~m\,s$^{-1}$ and MAE$=0.0883$~m\,s$^{-1}$.

The performance of OcDiffSR for the meridional component $(\mathbf{v})$ is comparable to the zonal velocity $(\mathbf{u})$, with RMSE$=0.046$~m\,s$^{-1}$ and MAE$=0.020$~m\,s$^{-1}$ slightly higher, indicating modestly greater reconstruction error. The PCC$=0.636$ and $R^2=0.374$ suggest moderate to poor correlation strength. SSIM$=0.694$ indicates moderate structural preservation of meridional flow features. PSNR$=25.2$~dB and NRMSE$=0.055$ indicate weaker performance compared to scalar variables. Similar to the zonal velocity, the evaluation metrics for meridional velocity $(\mathbf{v})$, predicted by the interpolation, perform better than those of CorrDiff, i.e., CorrDiff predicted values are RMSE$=0.1205$~m\,s$^{-1}$, MAE$=0.0524$~m\,s$^{-1}$, PCC$=0.546$, SSIM$=0.617$, and the interpolation model values for meridional velocity are RMSE$=0.0590$~m\,s$^{-1}$, MAE$=0.0275$~m\,s$^{-1}$, PCC$=0.3456$, SSIM$=0.618$. These results confirm that velocity reconstruction is inherently more challenging than scalar field reconstruction, likely owing to the anisotropic dynamics and high spatial intermittency of oceanic vector fields \cite{cheng2023machine}. Nevertheless, OcDiffSR better preserves dominant flow directions and mesoscale circulation structures than either baseline.

Overall, OcDiffSR outperforms both the interpolation and the CorrDiff model \cite{mardani2025}, demonstrating strong generalization and reconstruction capability, particularly in the case of the scalar oceanographic fields (sea surface temperature and sea surface salinity). In contrast, for velocity components $(u,v)$, while the error magnitudes remain small, the correlation coefficients PCC$\approx 0.64$ and SSIM$\approx 0.69$ suggest that improvement is needed in reproducing fine-scale vector flow structures. These differences likely arise from the more nonlinear and turbulent nature of oceanic velocity fields compared to scalar tracers.

\begin{figure}[ht]
    \centering
    \includegraphics[width=0.99 \textwidth]{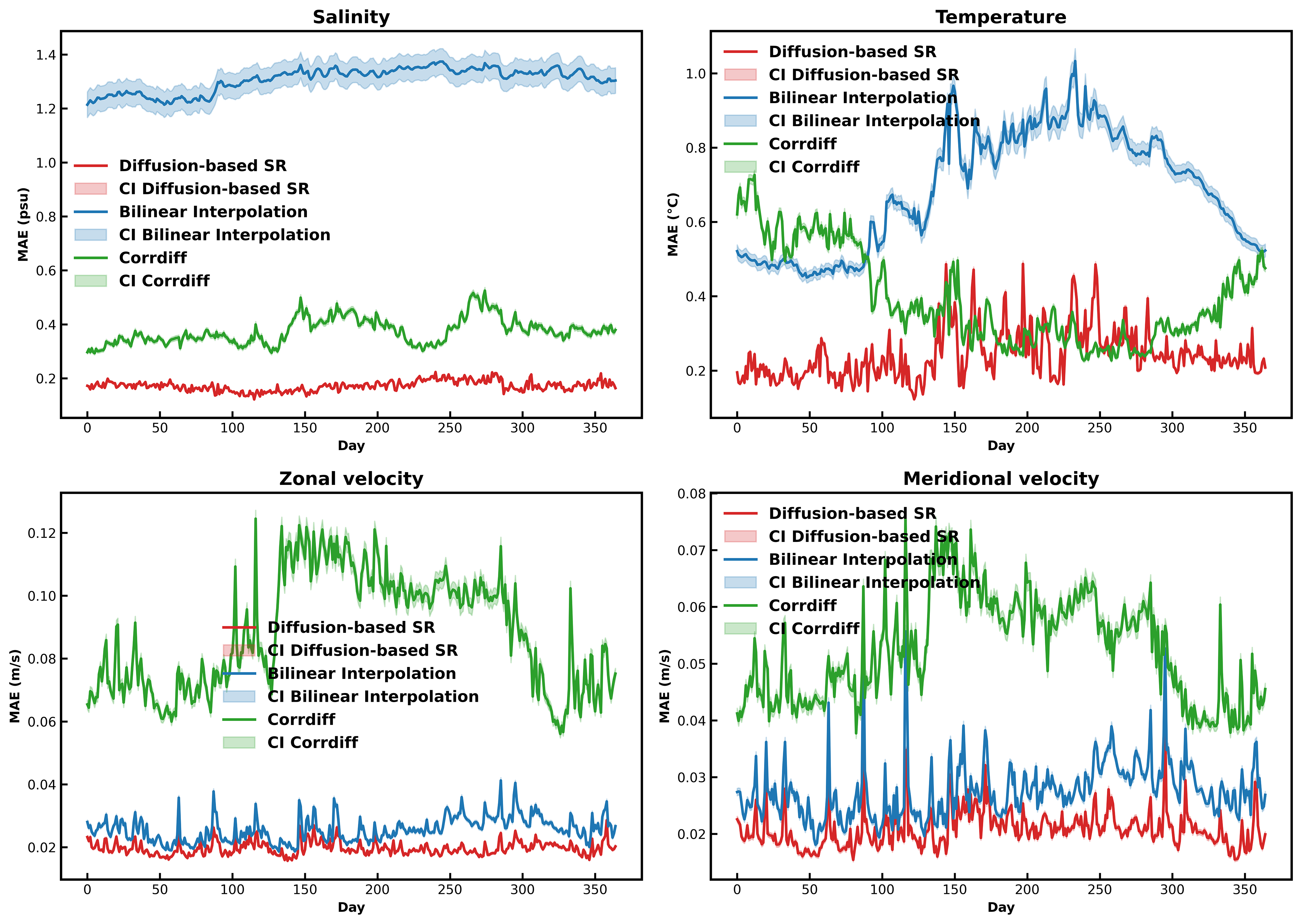}
    \caption{Daily Mean Absolute Error (MAE) comparison between OcDiffSR, bilinear interpolation, and Corrdiff model for four key oceanographic variables: salinity, temperature, zonal velocity, and meridional velocity.}
    \label{fig:all_variable_mae_ci}
\end{figure}

Figure~\ref{fig:all_variable_mae_ci} presents the daily MAE for all four oceanographic variables throughout the test year, together with 95\% confidence intervals (CIs), providing insight into the temporal stability and reliability of the methods.

OcDiffSR maintains consistently lower salinity MAE$\approx 0.15$-$0.20$~psu throughout the year. This indicates a stable and accurate reconstruction of salinity fields with minimal seasonal sensitivity. In contrast, bilinear interpolation exhibits errors roughly six to eight times higher, with MAE values increasing from $1.2-1.4$ psu over the whole year. The narrow CI band of the OcDiffSR model indicates stable performance with minimal temporal variability. The CorrDiff model shows a clear improvement over bilinear interpolation, reducing the MAE to approximately $0.30$--$0.45$~psu; however, its mean absolute error remains higher than that of the OcDiffSR model, which indicates larger temporal fluctuations, particularly during mid-year periods.

A similar pattern holds for SST: bilinear interpolation exhibits strongly seasonally dependent MAE, exceeding $1^{\circ}$C during summer (approximately days~150--250, corresponding to late May through early September). In the case of OcDiffSR, MAE maintains lower and more stable errors around $0.2-0.45 \degree $~C, with a narrow CI envelope across the entire year. The CorrDiff model performs better than bilinear interpolation with MAE approximately $0.25-0.75 \degree $~C. Nevertheless, CorrDiff exhibits noticeable seasonal variability and wider confidence intervals compared to OcDiffSR, suggesting comparatively lower temporal stability.

In the case of zonal velocity, OcDiffSR performs better than the bilinear interpolation, achieving MAE values between $0.016-0.022$~m\,s$^{-1}$, compared to $0.025-0.040 $~m\,s$^{-1}$ for interpolation. It can be seen that OcDiffSR suppresses the large error spikes observed in bilinear interpolation, indicating improved reconstruction of dynamic velocity structures. The small CI envelope further reflects higher temporal consistency and reduced uncertainty. The CorrDiff model performance is inferior compared to both the interpolation and OcDiffSR models, with MAE values typically in the range of $0.07$-$0.12 $~m\,s$^{-1}$. The CorrDiff model exhibits larger temporal variability and a broader CI envelope, particularly during summer periods.

The meridional velocity shows similar results to the zonal velocity $(\mathbf{u})$. OcDiffSR maintains low MAE values around $0.015$-$0.035 $~m\,s$^{-1}$, outperforming bilinear interpolation, which ranges between $0.02$ and $0.04 $~m\,s$^{-1}$ and has frequent high-amplitude peaks. Although both methods experience occasional peaks associated with energetic flow events, the OcDiffSR model effectively mitigates these extremes, resulting in smoother temporal error evolution and narrower confidence intervals. The CorrDiff model again demonstrates inferior performance over both the interpolation and OcDiffSR models in the case of meridional velocity, with MAE values around $0.04-0.07 $~m\,s$^{-1}$.

\begin{figure}[ht]
    \centering
    \includegraphics[width=0.95 \textwidth]{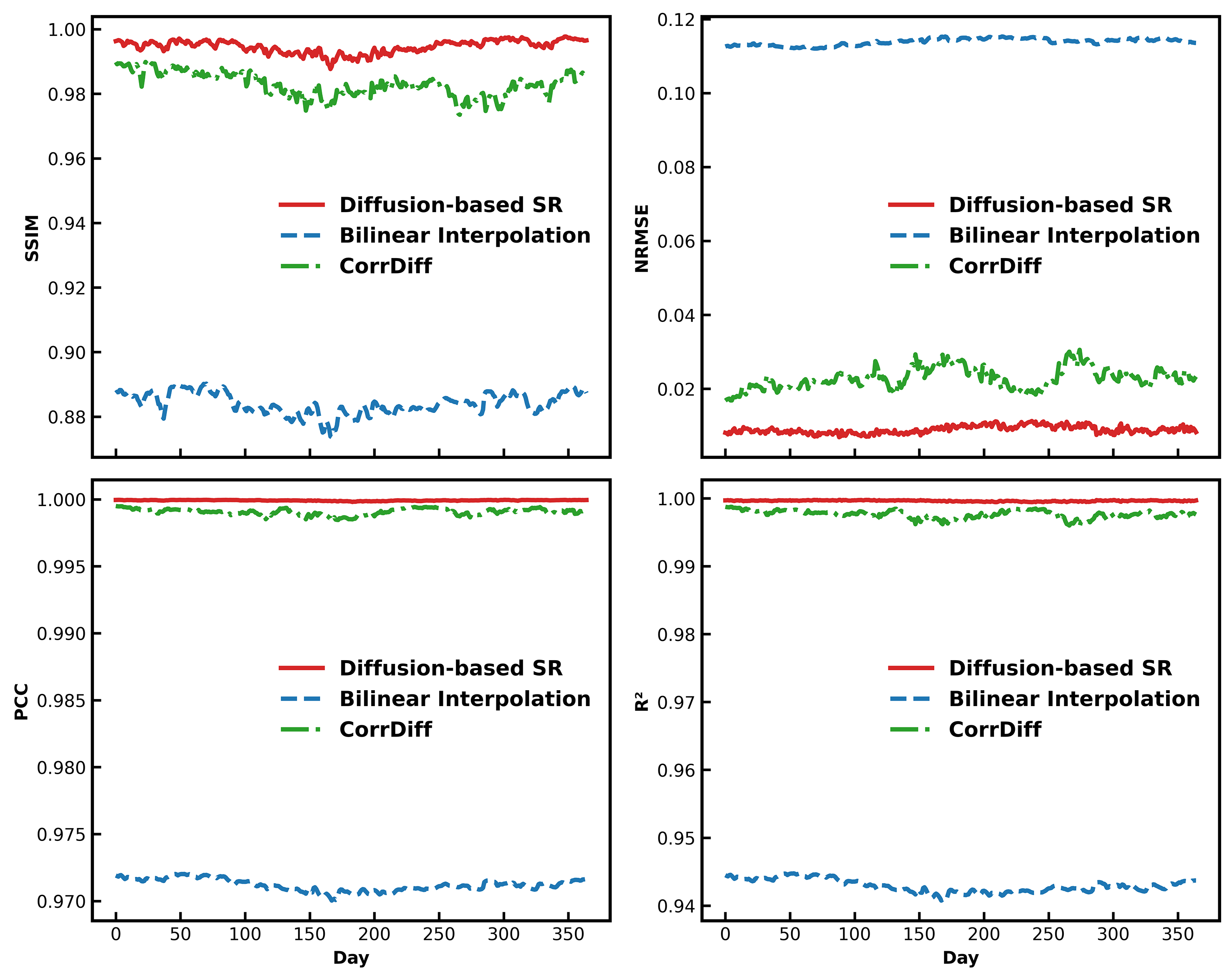}
    \caption{Comparison of four evaluation metrics (SSIM, NRMSE, PCC, and $R^2$) between OcDiffSR, bilinear interpolation, and CorrDiff model for daily salinity data.}
    \label{fig:metric_salinity_daily}
\end{figure}

\begin{figure}[ht]
    \centering
    \includegraphics[width=0.95 \textwidth]{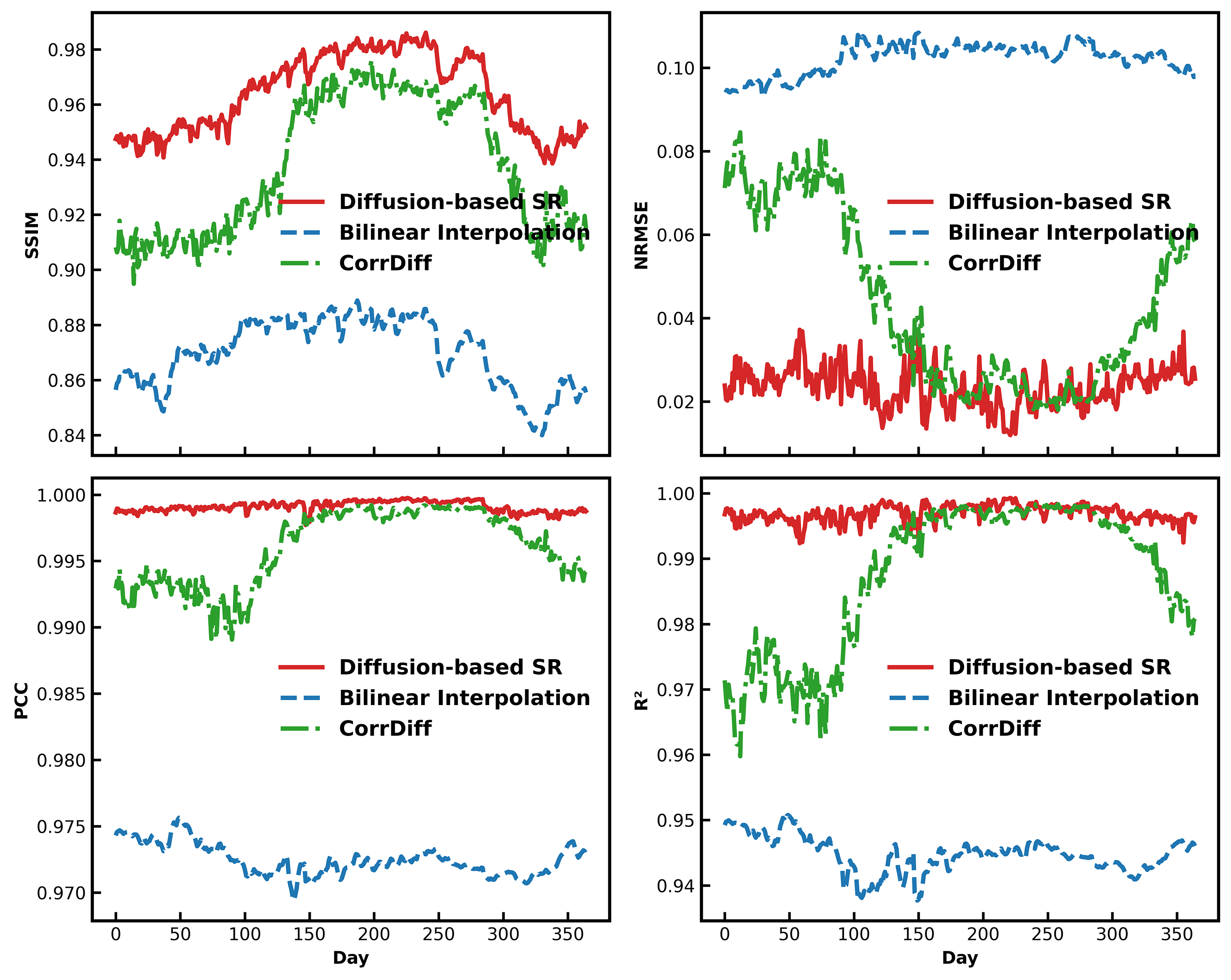}
    \caption{Comparison of four evaluation metrics (SSIM, NRMSE, PCC, and $R^2$) between OcDiffSR, bilinear interpolation, and CorrDiff model for daily sea surface temperature.}
    \label{fig:metric_temperature_daily}
\end{figure}

\begin{figure}[ht]
    \centering
    \includegraphics[width=0.95 \textwidth]{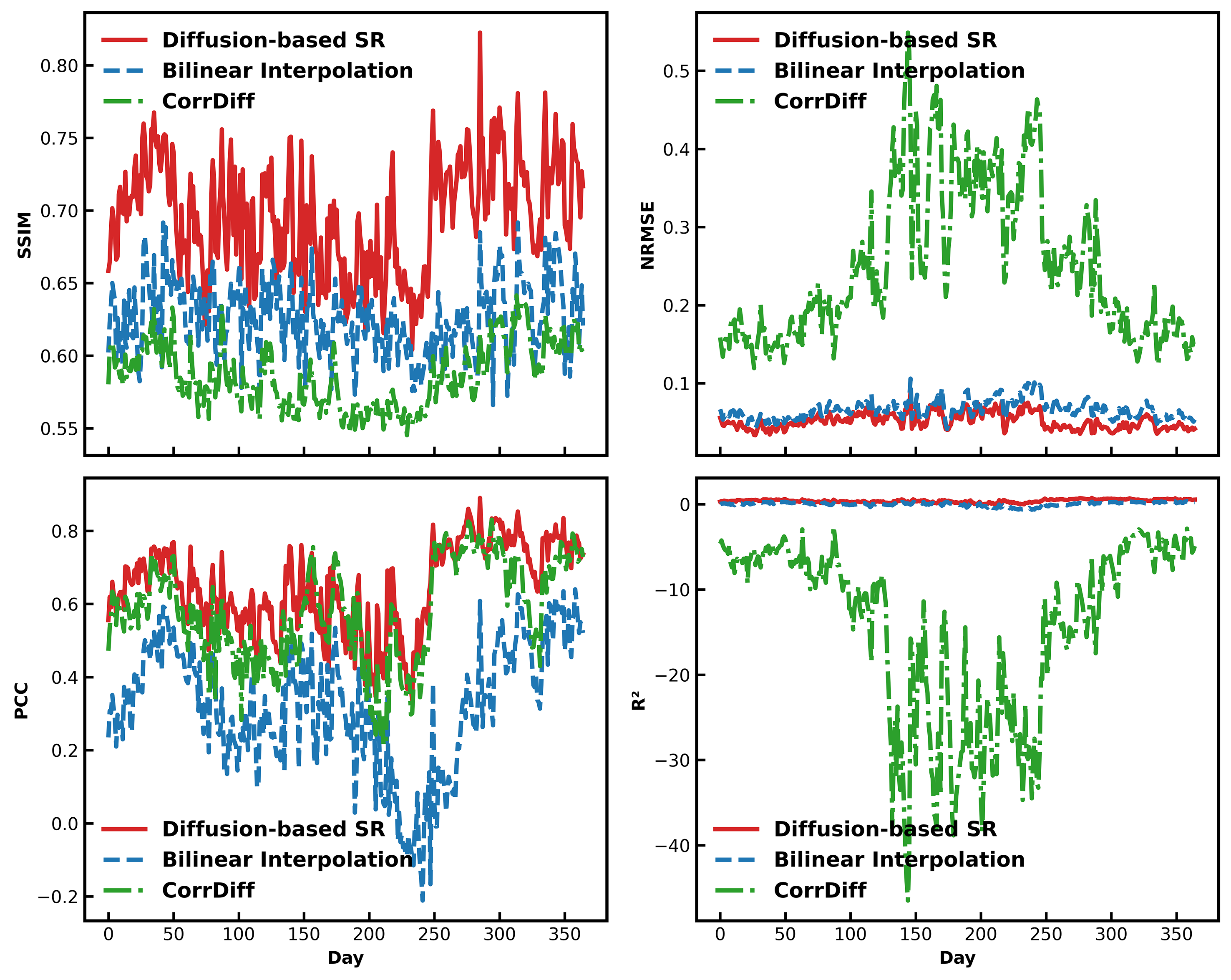}
    \caption{Comparison of four evaluation metrics (SSIM, NRMSE, PCC, and $R^2$) between OcDiffSR, bilinear interpolation, and CorrDiff model for daily zonal velocities.}
    \label{fig:metric_zonal_daily}
\end{figure}

\begin{figure}[ht]
    \centering
    \includegraphics[width=0.95 \textwidth]{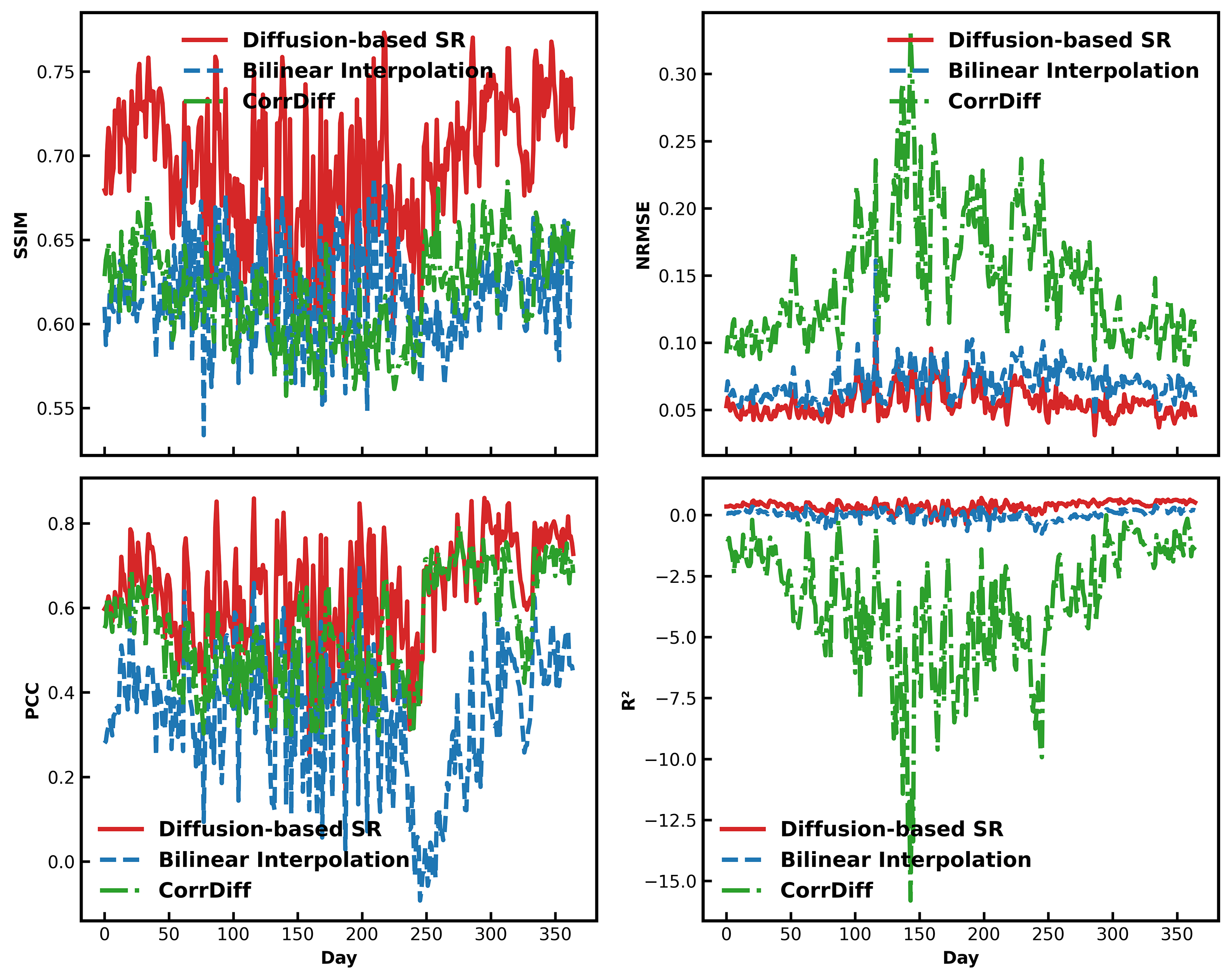}
    \caption{Comparison of four evaluation metrics (SSIM, NRMSE, PCC, and $R^2$) between OcDiffSR, bilinear interpolation, and CorrDiff model for daily meridional velocities.}
    \label{fig:metric_meridional_daily}
\end{figure}

We also evaluated the performance of the OcDiffSR model against conventional bilinear interpolation and the reference CorrDiff model using four standard metrics—SSIM, PCC, NRMSE, and $R^2$ computed daily for the entire year of 2009. These metrics jointly assess structural fidelity, spatial pattern agreement, relative error magnitude, and variance explained with respect to the high-resolution reanalysis dataset. SSIM is highlighted as the primary indicator of mesoscale structural preservation, while PCC and $R^2$ assess dynamical consistency and variance explanation. NRMSE provides a scale-independent measure of reconstruction error in different oceanographic variables. Figures \ref{fig:metric_salinity_daily}-\ref{fig:metric_meridional_daily} illustrate the temporal evolution of these metrics for the four oceanographic variables: salinity, sea surface temperature, zonal velocities, and meridional velocities.

For salinity (Figure \ref{fig:metric_salinity_daily}), OcDiffSR consistently achieves SSIM values near unity ($0.99-1.0$) throughout the year. Bilinear interpolation yields substantially lower SSIM ($\approx 0.88$-$0.89$), consistent with the oversmoothing of spatial features and loss of fine-scale structure inherent to kernel-based upsampling.  However, the CorrDiff model performs better than bilinear interpolation, with SSIM$\approx 0.98$. 

OcDiffSR achieves the lowest normalised errors (NRMSE~$< 0.01$) with only minor seasonal fluctuations. The CorrDiff model shows intermediate performance, with NRMSE values in the range $(0.02-0.03)$, and with temporal fluctuations, particularly during mid-year periods. Bilinear interpolation has larger errors with NRMSE$\approx 0.11$, showing its inability to recover accurate amplitudes at finer scales.

The Pearson correlation coefficient (PCC) of OcDiffSR remains close to unity ($\approx 0.999 -1.0$) throughout the year, indicating excellent spatial alignment between the predicted and reference salinity fields. CorrDiff also achieves high PCC values ($\approx 0.998-0.999$), which is slightly lower than that of OcDiffSR, with slightly larger temporal variability. Bilinear interpolation shows a noticeably lower correlation ($\approx 0.972$).

$R^2$ values from OcDiffSR are near unity throughout the year, indicating that the model explains nearly all variance in the reference reanalysis with negligible temporal degradation. CorrDiff explains slightly less variance, with $R^2$ values marginally below unity and small seasonal declines, indicating a partial loss of variability at certain times of the year. Bilinear interpolation shows lower values $R^2\approx 0.94$, representing a large fraction of salinity variability that is not recovered. Overall, these results emphasize that OcDiffSR not only reduces errors but also preserves the correct spatial phase of salinity variability, and its performance is superior to that of CorrDiff and interpolation.

Figure \ref{fig:metric_temperature_daily} presents the evaluation metrics predicted by OcDiffSR, interpolation, and CorrDiff in the case of sea surface temperature. SSIM values for sea surface temperature obtained from OcDiffSR are in the range $ 0.94-0.98$, which indicates the model's capability to preserve temperature gradients and spatial structures. A seasonal peak in SSIM is observed during mid-year, coinciding with the period of strongest thermal gradients and most spatially coherent SST patterns in the Adriatic Sea. Bilinear interpolation shows lower SSIM values ($\approx 0.86$), representing smoothing of thermal fronts and a reduced representation of mesoscale features. The CorrDiff model performs better than bilinear interpolation, with SSIM values in the range  $0.90-0.97$, particularly during mid-year. However, CorrDiff shows degradation during winter and transition seasons, indicating reduced robustness under varying thermal regimes. 

The NRMSE panel shows that the OcDiffSR model maintains low normalized errors, between $0.015$ and $0.04$. The CorrDiff shows intermediate performance, with NRMSE values decreasing to $0.02$ during mid-year but increasing to greater than $0.06$ during winter. Bilinear interpolation consistently exhibits the highest NRMSE values ($\approx 0.10$), demonstrating its limited capability to accurately reconstruct SST amplitudes across all seasons.

Further, Figure \ref{fig:metric_temperature_daily} shows that OcDiffSR achieves close to perfect correlation with the reference reanalysis, with PCC values remaining above $0.998$ throughout the year. This indicates excellent spatial alignment of temperature. The CorrDiff also demonstrates high PCC values ($\approx 0.998$), particularly during mid-year, but it exhibits noticeable declines during winter and transition periods. Bilinear interpolation shows the lowest correlation ($\approx 0.975$), confirming that while large-scale patterns are preserved, finer-scale spatial alignment is degraded.

Also, $R^2$ values achieved by OcDiffSR are close to $1$, with slight temporal variability, indicating that both large-scale seasonal cycles and mesoscale variability are effectively captured. The CorrDiff shows substantial seasonal dependence, with $R^2$ increasing to near-unity during mid-year but dropping to $\approx 0.96$ during winter periods. Bilinear interpolation yields the lowest $R^2$ values ($\approx 0.94$), suggesting a persistent loss of variance across all seasons. Overall, the OcDiffSR model provides the most consistent and structurally accurate SST reconstructions.

For zonal velocities (Figure \ref{fig:metric_zonal_daily}), the OcDiffSR model again provided superior performance compared to bilinear interpolation and the CorrDiff model. The OcDiffSR model achieves SSIM values ranging between $0.65$ and $0.80$, with noticeable temporal variability. The moderate SSIM values reflect the inherent difficulty of reconstructing fine-scale velocity gradients and coherent current structures at this spatial resolution. However, the SSIM values of OcDiffSR are higher than those obtained from the bilinear interpolation ($\approx 0.63$) and CorrDiff model ($\approx 0.6$) with the lowest SSIM values. 

The NRMSE results further emphasize the robustness of OcDiffSR. Normalized errors remain relatively low ($\approx 0.05$) compared to that of interpolation ($\approx 0.08$) and CorrDiff (in the range $0.15 - 0.5$).

The PCC panel reveals that OcDiffSR achieves low-to-moderate correlations $(\approx 0.45-0.85)$, but it still maintains positive and physically meaningful correlations. However, bilinear interpolation exhibits weaker and highly unstable correlations, including periods of near-zero or even negative PCC, indicating incorrect spatial alignment and phase errors in velocity anomalies. CorrDiff performs slightly better than bilinear interpolation but underperforms in comparison to the OcDiffSR model.

The $R^2$ values obtained by OcDiffSR for zonal velocities are low, i.e., near-zero but still positive throughout the year, indicating that it captures a meaningful fraction of the velocity variance without catastrophic failure. Similarly, bilinear interpolation exhibits near-zero $R^2$ values, reflecting limited but stable variance representation. In contrast, CorrDiff produces strongly negative $R^2$ values, particularly during dynamically active periods of mid-year. Such negative values indicate that CorrDiff performs substantially worse than a climatological mean predictor, failing to represent the dominant variability of the zonal velocity field.

Lastly, meridional velocities (Figure \ref{fig:metric_meridional_daily}) exhibit trends that closely mirror those of the zonal component. The SSIM values obtained from OcDiffSR for meridional velocity throughout the year range between $0.6$ and $0.75$, showing moderate preservation of meridional flow structures with noticeable temporal variability reflecting the dynamic nature of meridional currents, which are hard to capture by OcDiffSR. Bilinear interpolation shows lower SSIM values ($\approx 0.55 - 0.70$) indicating smoothing of velocity gradients. CorrDiff yields SSIM values ($\approx 0.58$-$0.68$) comparable to those of bilinear interpolation, with pronounced temporal fluctuations.

The NRMSE panel shows that OcDiffSR maintains relatively low normalized errors $(\approx 0.04-0.07)$ throughout the year, with only modest seasonal fluctuation. Bilinear interpolation produces higher NRMSE values $(\approx 0.06-0.09)$ in comparison to OcDiffSR, while the CorrDiff model performs worse with significantly elevated and highly variable errors, and NRMSE frequently exceeds $0.10$, peaking above $0.30$ during mid-year.

In the case of meridional velocity, the OcDiffSR model shows low to moderate PCC values (in the range $0.3-0.8$), indicating partial alignment of meridional velocity relative to the reference reanalysis data. However, the bilinear interpolation exhibits weaker and more unstable correlations, including intervals of near-zero and occasionally negative PCC, reflecting frequent phase errors and incorrect anomaly placement. The CorrDiff performs better than bilinear interpolation during some periods but still shows pronounced temporal instability and reduced correlations during dynamically active phases. 

The $R^2$ values achieved by OcDiffSR are above zero across most of the year, indicating that it captures a meaningful fraction of meridional velocity variance without severe degradation. The bilinear interpolation exhibits similar but slightly more variable behavior, with small positive or near-zero $R^2$ values. In contrast, CorrDiff gives negative $R^2$ values, which indicate worse performance than the mean predictor.

\begin{figure}[ht]
    \centering
    \includegraphics[width=0.95 \textwidth]{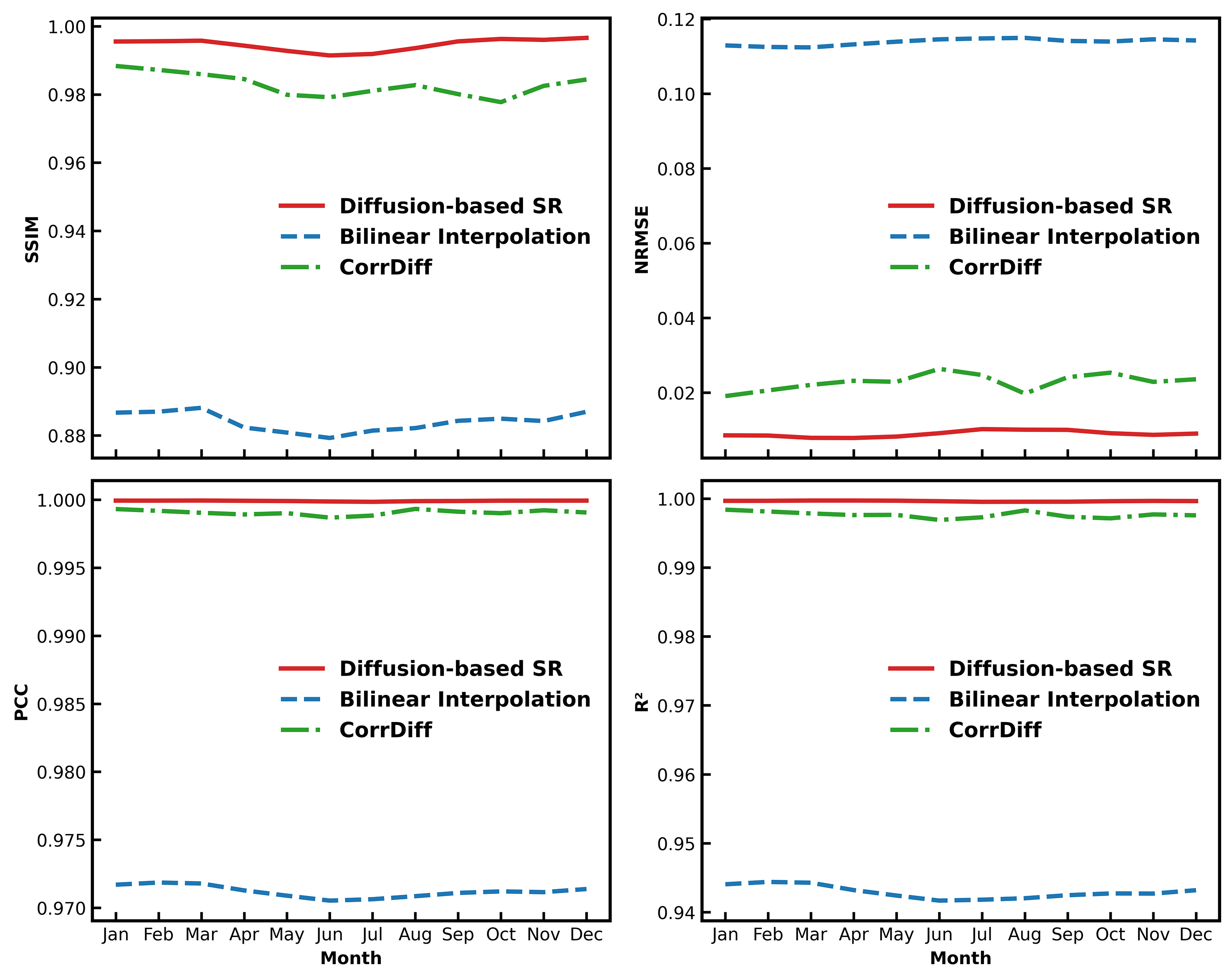}
    \caption{Comparison of four evaluation metrics (SSIM, NRMSE, PCC, and $R^2$) between OcDiffSR, bilinear interpolation, and CorrDiff model for monthly salinity data.}
    \label{fig:metric_salinity_monthly}
\end{figure}

\begin{figure}[ht]
    \centering
    \includegraphics[width=0.95 \textwidth]{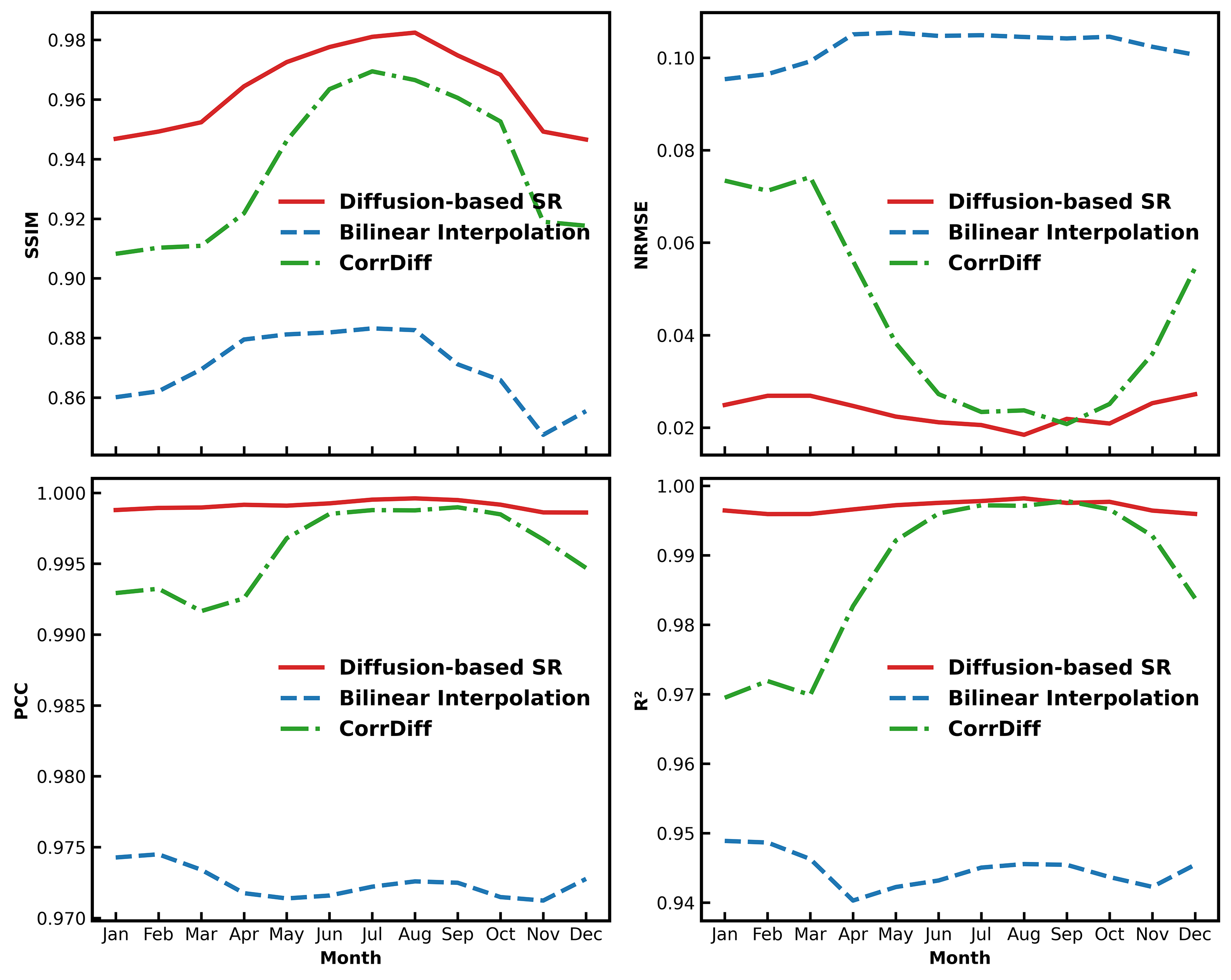}
    \caption{Comparison of four evaluation metrics (SSIM, NRMSE, PCC, and $R^2$) between OcDiffSR, bilinear interpolation, and CorrDiff model for monthly sea surface temperature.}
    \label{fig:metric_temperature_monthly}
\end{figure}

\begin{figure}[ht]
    \centering
    \includegraphics[width=0.95 \textwidth]{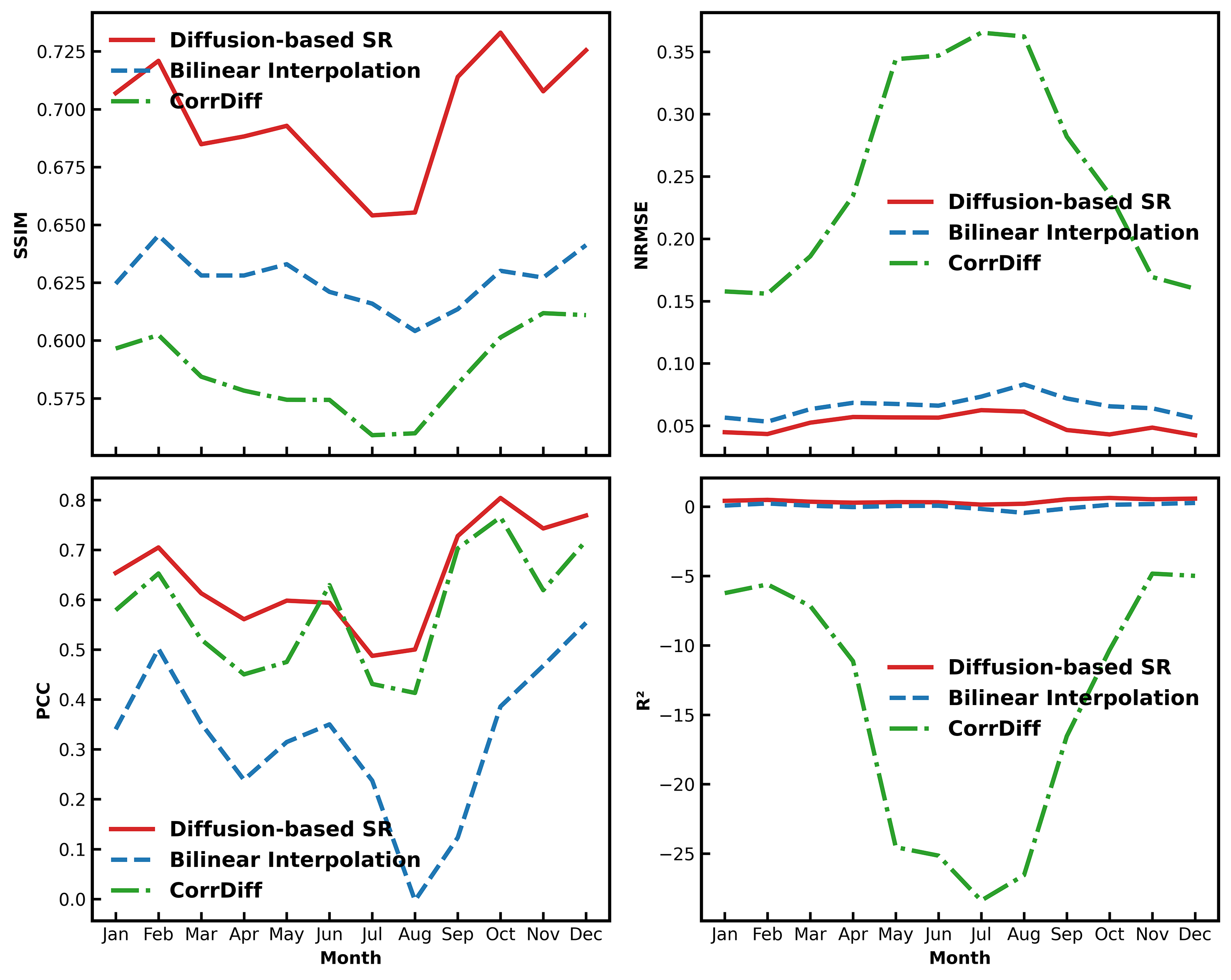}
    \caption{Comparison of four evaluation metrics (SSIM, NRMSE, PCC, and $R^2$) between OcDiffSR, bilinear interpolation, and CorrDiff model for monthly zonal velocities.}
    \label{fig:metric_zonal_monthly}
\end{figure}

\begin{figure}[ht]
    \centering
    \includegraphics[width=0.95 \textwidth]{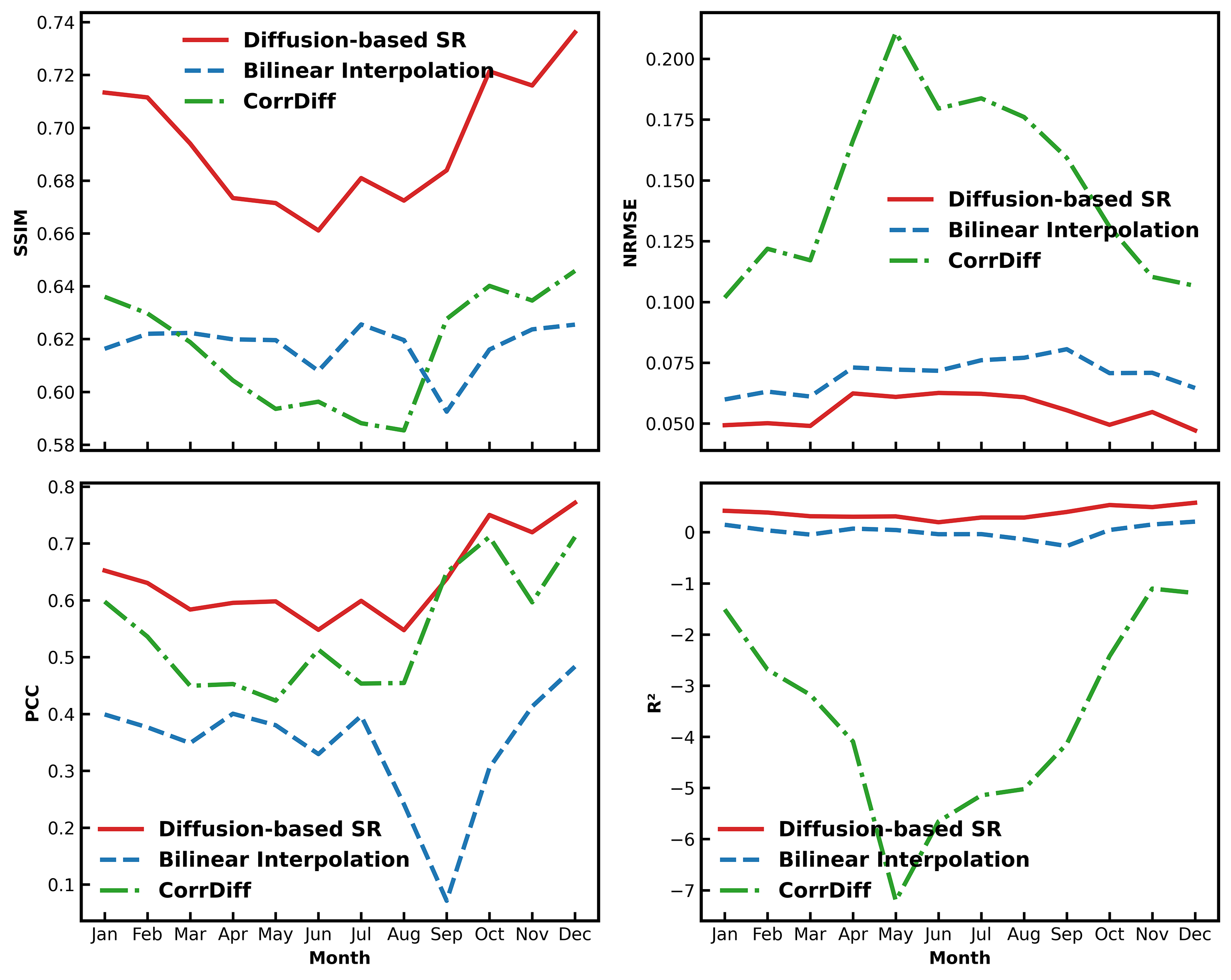}
    \caption{Comparison of four evaluation metrics (SSIM, NRMSE, PCC, and $R^2$) between OcDiffSR, bilinear interpolation, and CorrDiff model for monthly meridional velocities.}
    \label{fig:metric_meridional_monthly}
\end{figure}

Further, to assess the model's performance, we computed the monthly aggregated evaluation metrics, which reduce the short-term variability and highlight the seasonal consistency and long-term robustness of the model, and presented them in Figures \ref{fig:metric_salinity_monthly} - \ref{fig:metric_meridional_monthly}. 

The monthly evaluation metrics of salinity (Figure \ref{fig:metric_salinity_monthly}) reveal that OcDiffSR achieves near-perfect structural similarity (SSIM$ >0.99$), spatial coherence PCC close to unity, and minimal normalized error (NRMSE$ <0.01$), indicating stable and accurate reconstruction of large-scale and mesoscale salinity patterns in comparison with the high-resolution reanalysis data across all months of the year 2009. CorrDiff exhibits slightly reduced SSIM ($\approx 0.98$) and increased seasonal variability in NRMSE ($\approx 0.02-0.03$), particularly during late spring and summer, while bilinear interpolation shows substantially lower SSIM ($\approx 0.88$), higher NRMSE ($\approx 0.11$), and weaker spatial correlation (PCC$ \approx 0.97$). Consistent with the above trends, the OcDiffSR model explains nearly all variance in the reference field ($R^2 \approx 1$), outperforming CorrDiff and bilinear interpolation across seasons.

As shown in Figure \ref{fig:metric_temperature_monthly}, the OcDiffSR model substantially improves the temperature field reconstruction over CorrDiff and bilinear interpolation. OcDiffSR achieves SSIM$\approx 0.96$ across all months, indicating superior preservation of spatial structures and thermal fronts relative to bilinear interpolation (SSIM$ \approx 0.86$) and CorrDiff (SSIM$ \approx 0.93$). Error-based evaluation shows that the OcDiffSR model maintains the lowest NRMSE$\approx 0.02$ throughout the year, whereas bilinear interpolation exhibits persistently higher errors (NRMSE$ \approx 0.09$) and limited seasonal sensitivity. However, CorrDiff demonstrates improved performance during mid-year months (NRMSE$\approx 0.03$) but degrades toward winter ($\approx 0.04-0.08$), suggesting reduced robustness under weaker thermal gradients. Correlation metrics further highlight the advantage of the OcDiffSR model, with near-unity PCC and $R^2$ values across all months, reflecting stable temporal coherence and variance preservation. In contrast, bilinear interpolation systematically underestimates variability ($R^2\approx 0.94$), while CorrDiff exhibits seasonal instability ($R^2\approx 1$ in summer, while $R^2$ ranges between $0.97-0.99$ in winter). 

Figure \ref{fig:metric_zonal_monthly} presents the monthly evaluation metrics for zonal velocities.

The OcDiffSR model shows the highest SSIM (in the range $\approx 0.65-0.72$) with seasonal fluctuations compared to bilinear interpolation and CorrDiff. In terms of accuracy, OcDiffSR maintains the lowest NRMSE ($\approx 0.05$) across all months, while bilinear interpolation shows a slightly higher NRMSE than OcDiffSR. The CorrDiff exhibits the highest error (NRMSE ranging in $0.15-0.35$) with a pronounced mid-year error amplification, suggesting sensitivity to intensified summer circulation. Correlation-based metrics show that OcDiffSR achieves higher PCC values, reflecting reliable temporal coherence, whereas bilinear interpolation shows weak and intermittent correlations. Notably, CorrDiff yields strongly negative $R^2$ values during summer months, indicating a failure to reproduce variance in periods of enhanced dynamical complexity.

Figure \ref{fig:metric_meridional_monthly} illustrates the monthly performance of the OcDiffSR model, the bilinear interpolation, and the CorrDiff model for meridional velocity. Across all months, OcDiffSR achieves the highest SSIM (in the range $\approx 0.67-0.74$) relative to bilinear interpolation and CorrDiff. OcDiffSR shows its weakest reconstruction performance during summer, when meridional flow structures are most energetic. OcDiffSR maintains the lowest NRMSE ($\approx 0.06$) throughout the year, whereas CorrDiff exhibits substantially elevated errors during late spring and summer, coinciding with periods of enhanced meridional variability. Correlation analysis shows that OcDiffSR consistently yields higher PCC values, while bilinear interpolation demonstrates weak correlations and pronounced degradation during transitional months. Notably, CorrDiff produces strongly negative $R^2$ values over most of the year, indicating an inability to reproduce variance in meridional currents.  

Across all evaluated variables, the OcDiffSR model consistently outperforms bilinear interpolation and the CorrDiff model in daily and monthly metrics. OcDiffSR achieves lower reconstruction errors and higher statistical consistency, demonstrating its robustness and adaptability to diverse oceanic properties. These results confirm that OcDiffSR effectively leverages the spatial context provided by the low-resolution conditioning input, enabling realistic high-resolution reconstructions that maintain dynamical coherence across all variables and seasons. Its performance across both scalar (salinity, temperature) and vector (velocity) variables shows its potential as a generalized super-resolution tool for ocean reanalysis and model downscaling applications.

\section{Conclusion}
This study has introduced OcDiffSR, a conditional denoising diffusion probabilistic model for oceanographic super-resolution in the Adriatic Sea. Trained on a decade of paired GLORYS12V1 and Med~MFC reanalysis data and evaluated on the independent test year 2009, the framework integrates multi-scale low-resolution conditioning, cross-attention bottleneck layers, and sinusoidal seasonal embeddings via FiLM to jointly reconstruct sea-surface temperature, salinity, and horizontal velocity components. The temporal split, in which the test year predates the training period, ensures strict independence from reanalysis autocorrelation and provides a rigorous basis for generalisation assessment.

For scalar oceanographic fields, OcDiffSR consistently outperforms both bilinear interpolation and the reference CorrDiff model across all evaluated metrics. The model achieves near-perfect spatial correlation (PCC~$\geq 0.999$), near-unity coefficient of determination ($R^2 \geq 0.997$), and high structural similarity (SSIM~$\geq 0.964$) for sea-surface salinity and temperature, with RMSE values of $0.346$~psu and $0.477\degree$C respectively. Daily and monthly evaluations confirm temporal stability across all seasons, with narrow confidence intervals and no significant seasonal degradation. These results demonstrate the model's ability to recover fine-scale gradients, mesoscale structures, and coherent circulation patterns that are absent in low-resolution inputs and inadequately captured by conventional downscaling methods.
For vector fields, OcDiffSR achieves the lowest absolute errors among all three methods (RMSE~$= 0.044$~m\,s$^{-1}$ for zonal velocity), preserves dominant flow directions and mesoscale circulation structures, and suppresses the large error spikes and negative $R^2$ values exhibited by CorrDiff during dynamically active periods. However, moderate correlation coefficients (PCC~$\approx 0.64$) and structural similarity (SSIM~$\approx 0.69$) reflect the intrinsic difficulty of reconstructing highly intermittent and anisotropic oceanic velocity fields, and indicate that further improvement is needed for operational vector field downscaling.

Future work will address two principal directions. First, physics-informed regularisation, such as geostrophic balance constraints or divergence-free penalties, will be incorporated to improve the dynamical consistency of reconstructed velocity fields, complemented by quantitative physical-consistency diagnostics (e.g.\ divergence, vorticity, or power-spectral-density analysis), since the present study assesses physical plausibility only qualitatively through visual inspection of the reconstructed fields. Second, while OcDiffSR natively supports ensemble-based uncertainty quantification through repeated stochastic sampling of the reverse diffusion process, a rigorous empirical characterisation of ensemble spread, spread-error consistency, and calibration against independent observations is deferred to a dedicated subsequent study. Such analysis will further establish the operational readiness of the framework for probabilistic oceanographic downscaling, ensemble-based data assimilation, and transferability to other semi-enclosed basins and three-dimensional subsurface field estimation. Beyond these two directions, systematic ablation studies isolating the contribution of individual architectural components, such as the multi-scale low-resolution encoder, the FiLM-based seasonal conditioning, and joint versus per-variable training, were not performed in this study and are identified as a further priority for disentangling the sources of the reported performance gains. Additionally, the 365-day evaluation reported here characterises robustness across differing dynamical conditions (between-day variability) through descriptive statistics and 95\% confidence intervals, but it does not, by itself, quantify the stochastic sampling variance associated with repeated sampling of a fixed conditioning input (within-day variability, to be addressed by the ensemble-spread characterisation noted above), nor does it include formal statistical-significance testing, such as paired tests across the 365 daily values, between OcDiffSR and the baseline methods; such analysis is left to future work.

\section*{Acknowledgments}
RS acknowledges the CINECA award under the ISCRA initiative for the provision of high-performance computing resources and technical support.

\subsection*{Author Contributions} 

R. Srivastava: Conceptualization, Methodology, Model Development, Data Curation, Formal Analysis, Visualization, Writing—Original Draft, Writing—Reviewing and Editing.\\
M. Sarmad: Methodology, Model Development, Investigation, Writing—Reviewing and Editing.\\
E. Mele: Methodology, Model Development, Investigation, Writing—Reviewing and Editing.\\
M. Cafaro: Methodology, Supervision, Funding Acquisition, Writing—Review and Editing.\\
M. Pulimeno: Writing—Reviewing and Editing.\\
I. Epicoco: Conceptualization, Methodology, Supervision, Funding Acquisition, Writing—Review and Editing.\\
All authors approved the final version of the manuscript.

\subsection*{Funding}
This work was partially funded under the National Recovery and Resilience Plan (NRRP), Mission 4, Component 2, Investment 1.4- Call for Tender No. 1031 of 17/06/2022 by the Italian Ministry for University and Research, funded by the European Union- NextGenerationEU (Project No. CN-00000013).

\subsection*{Conflicts of Interest}
``The authors declare that there is no conflict of interest regarding the publication of this article.''

\subsection*{Data Availability}
The low-resolution forcing data used in this study are freely available from the Copernicus Marine Environment Monitoring Service (CMEMS) as the Global Ocean Physics Reanalysis GLORYS12V1 product (\url{https://doi.org/10.48670/moi-00021}). The high-resolution ground truth data are available as the Mediterranean Sea Physics Reanalysis (Med~MFC) product (\url{https://doi.org/10.25423/CMCC/MEDSEA\_MULTIYEAR\_PHY\_006\_004\_E3R1}). Both datasets are openly accessible upon free registration. The model code and training configuration used in this study are available from the corresponding author upon reasonable request.

\bibliography{diffusion_SR_bib}
\end{document}